\documentclass[aps,prl,twocolumn,superscriptaddress]{revtex4-1}
\usepackage{amsmath}
\usepackage{amssymb}
\usepackage{graphicx}
\usepackage{subfigure}
\usepackage{color}
\usepackage{float}
\usepackage{graphics}
\usepackage{hyperref}
\usepackage{txfonts}
\usepackage{tabularx}
\usepackage{array}

\begin{document}
\title{Topological magnons in a collinear altermagnet}

\author{Meng-Han Zhang}
\affiliation{Guangdong Provincial Key Laboratory of Magnetoelectric Physics and Devices, State Key Laboratory of Optoelectronic Materials and Technologies, Center for Neutron Science and Technology, School of Physics, Sun Yat-sen University, Guangzhou, 510275, China}

\author{Lu Xiao}
\affiliation{Guangdong Provincial Key Laboratory of Magnetoelectric Physics and Devices, State Key Laboratory of Optoelectronic Materials and Technologies, Center for Neutron Science and Technology, School of Physics, Sun Yat-sen University, Guangzhou, 510275, China}

\author{Dao-Xin Yao}
\email{yaodaox@mail.sysu.edu.cn}
\affiliation{Guangdong Provincial Key Laboratory of Magnetoelectric Physics and Devices, State Key Laboratory of Optoelectronic Materials and Technologies, Center for Neutron Science and Technology, School of Physics, Sun Yat-sen University, Guangzhou, 510275, China}

\begin{abstract}
We propose a model with Weyl magnons and nodal-line magnons ($\mathbb{Z}_2$) in a collinear altermagnet on the honeycomb lattice. Altermagnetic magnon bands can be realized by breaking the symmetry of the second nearest neighbor exchange couplings without the Dzyaloshinskii-Moriya interaction (DMI). In addition to the Chern number and $\mathbb{Z}_2$ invariant, chirality is important to describe the magnon topology. The model shows the nonzero magnon spin Nernst effect when a longitudinal temperature gradient exists. We calculate the differential gyromagnetic ratio induced purely by the topology of altermagnetic magnons, which can be probed through the Einstein-de Haas (EdH) effect. 
\end{abstract}
\date{\today}
\maketitle

\textit{\color{blue}Introduction.-} Chiral magnons in collinear altermagnets possess unique topological responses with symmetry-compensated zero net magnetization \cite{MnTe,RuO2,CrSb,ExperimentalMaterial}, facilitating spin transport via low-energy consumption \cite{MagnonHall,MagnonSpintronics,MagnonOrbitalNernst}. MnTe displays a spontaneous anomalous Hall effect (AHE) arising from hexagonal crystal altermagnetic order, which provides unique symmetry operations like inversion symmetry, mirror symmetry, and $C_{6v}$ rotational symmetry \cite{MnTe_1,MnTe_2}. The extraordinary spin splitting highlights g-wave altermagnetism characterized by the $\bar{6}^{2}/m^{2}m^{2}m^{1}$ spin Laue group, which incorporates symmetry transformations that decouple lattice and spin space \cite{Mn5Si3}. To avoid spin decoherence typically associated with relativistic spin-splitting phases, both MnSe and MnPSe$_{3}$ have been proposed to exhibit significant spin-momentum locking with i-wave altermagnetism \cite{MnSe}. The zero net magnetic moment and reduced stray fields in altermagnets can significantly improve the quality of magnet-superconductor heterostructures and magnet-semiconductor-superconductor wires, which are important in quantum computation \cite{AltermagneticMajorana,highordertopology}. Besides the solid state materials, the altermagnets can be simulated via atomic spins on surface using STM, ultracold atoms in optical lattices, Rydberg atoms, and topolectrical circuits \cite{Atomicspins,Coldatom,Rydbergcold,Rydbergcold1,TopolectricalCircuits}. 


\begin{figure}[t]
\centering
{
\includegraphics[width=\linewidth]{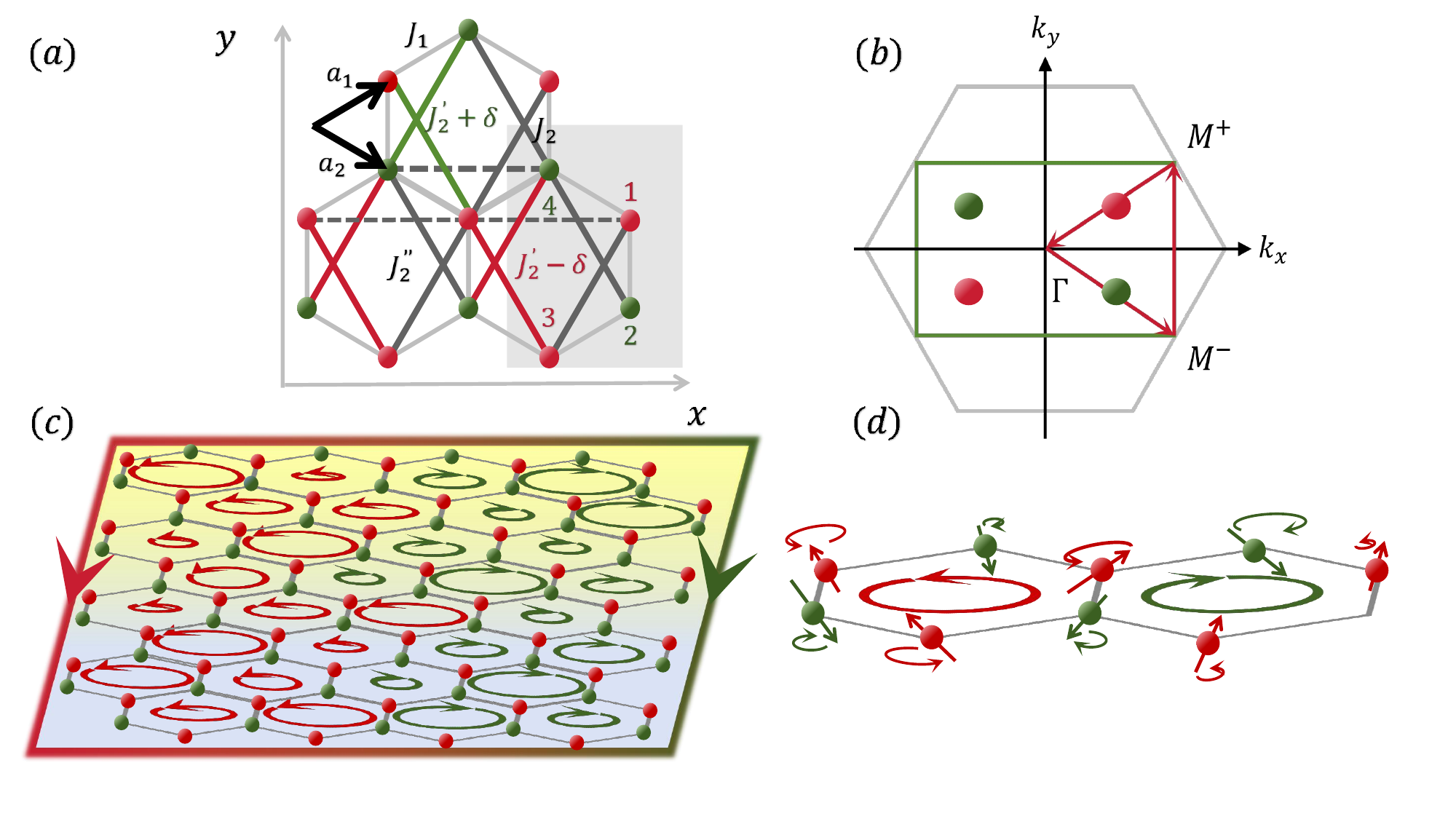}
}
\caption{(a) Altermagnet on the honeycomb lattice in magnetic N$\acute{e}$el order with shaded region that represents the unit cell. (b) Brillouin zones (BZ) for the lattice (grey) and magnetic structure (green). (c) Schematics of the magnon spin Nernst effect inducing magnon currents in transverse direction. (d) Magnon angular momentum consisting of left-hand (red) and right-hand (green) chirality.}
\label{Fig1}
\end{figure}

Although pairwise magnons with opposite chiralities occur in conventional antiferromagnets \cite{AntiSpintronics}, the degeneracy hinders a pure spin current mediated by magnons due to the inability to separate the spin states \cite{MagnonSNEinAnti1,MagnonSNEinAnti2}. However, altermagnetism is activated by non-relativistic spin splitting and hence reveals the lifting of Kramers degeneracy via the combined symmetry of mirror and time reversal \cite{MnTe_1,Landscape,SpinSpaceGroup}. Reflecting the spin-polarized band structure in altermagnets \cite{AltermagnonNernst}, the anisotropy of spin accumulation on chiral edge states has the potential to probe magnon-mediated spin currents and related magnon dynamics. Chiral nodal-lines evolve into pairs of Weyl magnons, providing valuable insight to understand and manipulate spin signals for magnon transport \cite{AlterWeyl}. For instance, non-zero Berry curvature endows the magnon wave package with angular momentum that can be decomposed into the self-rotation and the edge current \cite{OAM,Murakami1,Murakami2}. Relating the magnetic moment and angular momentum \cite{niexin1}, the gyromagnetic effect extends the understanding of magnetic properties into the realm of quasiparticles with exotic magnon topology \cite{LiJun,MHZhang1}. Demonstrating angular momentum textures and their subsequent conversion into mechanical rotation \cite{1915EdH,1908EdH,Barnett,EdH1}, the Einstein-de Haas (EdH) effect describes a unique gyromagnetic response triggered by altermagnetic magnons \cite{Gyratio,Landscape1,Landscape2}. Considering the non-relativistic symmetries of altermagnets, relevant measurements of EdH effect involve the topological Hall effect of magneton flow and magnon spin Nernst effect, which open up new possibilities for next-generation magnonic devices. 

\begin{figure*}[t]
\centering
{
\includegraphics[width=\linewidth]{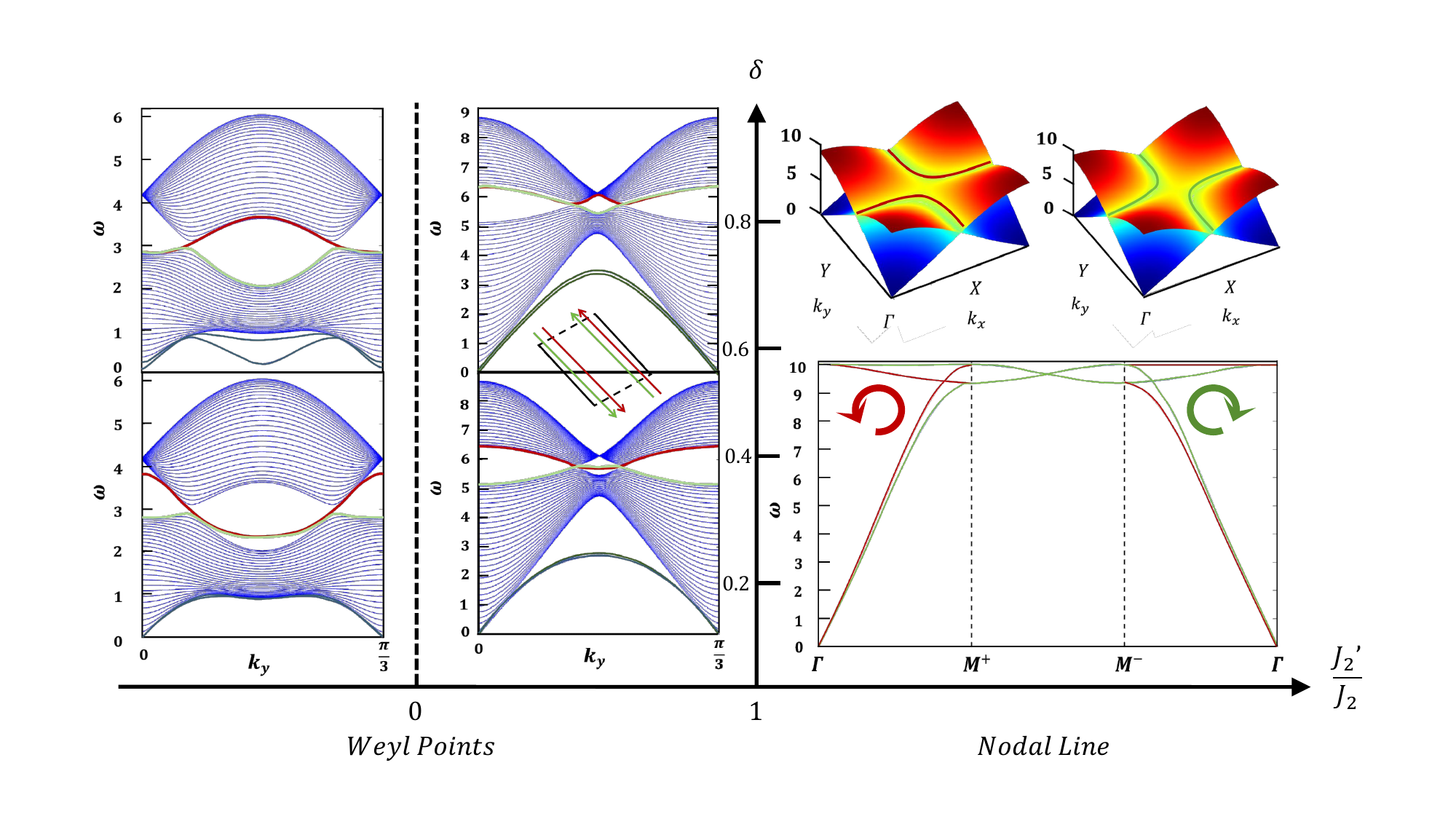}
}
\caption{Phase diagram of magnons in the ($J_2'/J_2$, $\delta$) plane. Weyl magnons and nodal-line magnons are separated by $J_2'/J_2 =1$ with chiralities indicated by the circle arrows and red/green color. The inset shows the magnon currents along the opposite chiral edges. The number of periodic 1D chains is 25.}
\label{Fig2}
\end{figure*}

In this Letter, we investigate the topological responses of magnon angular momenta unveiling the magnon transport properties of the honeycomb altermagnet in the absence of SOC \cite{SpontaneousAnomalousHall,RuO2_1}. While the  Dzyaloshinskii-Moriya interaction (DMI) contributes to spin accumulation by introducing chiral magnetic properties and spin textures \cite{DMI1,DMI2}, the efficiency can be limited by SOC and specific lattice geometry \cite{DMI3}. Compared to those systems dominated by the DMI, we propose that the combined symmetry operations in altermagnets enhance spin currents and accumulations through the alternating exchange couplings and topological magnons via chiral splitting \cite{AlterM1,Zeemansplitting}. We theoretically calculate the differential gyromagnetic ratios to understand the manifestation of topological EdH effect, which is beneficial for developing next-generation magnonic devices that are stable to external fields and impurities \cite{QuantumComputation,TopolectricalCircuits}. 

\textit{\color{blue}Model and Topology.-}
We construct a Heisenberg model on the honeycomb lattice with the N$\acute{e}$el order shown in Fig.~\ref{Fig1}

\begin{equation}
\begin{split}
H = J_{1}\sum_{\langle mn\rangle}\mathbf{S}_{m}\cdot \mathbf{S}_{n}-(J_2,J_{2}'\pm\delta, J_{2}'')\sum_{\langle \langle mn\rangle \rangle}\mathbf{S}_{m}\cdot \mathbf{S}_{n},
\end{split}
\end{equation} \label{Eq1}
where $J_{1}$$>$0 is the nearest neighbor coupling and $J_2$, $J_{2}'\pm\delta$ and $J_{2}''$ represent three types of next nearest neighbor (NNN) couplings, which host the altermagnetic magnon bands via breaking the symmetry of sublattices. The lattice vectors are defined as $\boldsymbol{a_1}$=$(\frac{\sqrt{3}}{2}, \frac{-1}{2})a$ and $\boldsymbol{a_2}$=$(\frac{\sqrt{3}}{2}, \frac{1}{2})a$. Using the Fourier transform, we find the kernel matrix $H=$$\sum_k \psi_{k}^{\dag} H_{k}\psi_{k}^{ }$ with

\begin{equation}
\begin{split}
H_k&=S\sigma_{0}\otimes[h_{0}\tau_{0}-f_{1}(\boldsymbol{k})\tau_{+}-f_{1}^{\dag}(\boldsymbol{k})\tau_{-}]\\
&-S\sigma_{+}\otimes[f_{2}(\boldsymbol{k})\tau_{+}+f_{2}^{\dag}(\boldsymbol{k})\tau_{-}+g_{\chi}(\boldsymbol{k})\tau^{0}_{+}+g_{\chi}(\boldsymbol{k})\tau^{0}_{-}]\\
&-S\sigma_{-}\otimes[f_{2}(\boldsymbol{k})\tau_{+}+f_{2}^{\dag}(\boldsymbol{k})\tau_{-}+g_{\chi}^{\dag}(\boldsymbol{k})\tau^{0}_{+}+g_{\chi}^{\dag}(\boldsymbol{k})\tau^{0}_{-}],
\end{split}\label{Eq2}
\end{equation}

where the space of the Hamiltonian is extended via the Nambu spinor $\psi_{k}^{\dag}$$\equiv$$[\alpha^{\dag}_{1,k}, \beta^{ }_{2,-k}, \alpha^{\dag}_{3,k}, \beta^{ }_{4,-k}]$, and the indices $\alpha$ and $\beta$ stand for bosonic operators with opposite spins. The altermagnetism is described by the generalized Bogoliubov-de Gennes (BdG) equation with the basis vectors $\sigma$ acting on the particle-hole space. Here, we use $\sigma_{\pm}(\tau_{\pm})$=$\frac{1}{2}[\sigma_{x}(\tau_{x})\pm i\sigma_{y}(\tau_{y})]$ and $\sigma^{0}_{\pm}(\tau^{0}_{\pm})$=$\frac{1}{2}[\sigma_{0}(\tau_{0})\pm \sigma_{z}(\tau_{z})]$ to analytically obtain the magnon bands
\begin{equation}
\begin{split}
\hbar\omega_{\chi}^{\pm}&=S\sqrt{\Lambda(\boldsymbol{k})\pm\lambda(\boldsymbol{k})},\\
\end{split} \label{Eq3}
\end{equation}

where the $+(-)$ sign corresponds to the acoustical (optical) branch, $\Lambda(\boldsymbol{k})$ and $\lambda(\boldsymbol{k})$ are related functions constructed by $h_{0}$, $f_{1}(\boldsymbol{k})$, $f_{2}(\boldsymbol{k})$ and $g_{\chi}(\boldsymbol{k})$ \cite{MHZhang3}. The altermagnetic magnons can be diagonalized via a Bogoliubov transformation $U$ from its eigenvectors $\Psi^{}_{\boldsymbol{k}}$=$U^{\dag}\psi^{}_{\boldsymbol{k}}$

\begin{equation}
\begin{split}
U &\equiv\sigma_{0} \otimes \left[
\begin{array}{cc}
\cosh{\frac{\vartheta}{2}} & -\sinh{\frac{\vartheta}{2}}e^{-i\varphi}\\
-\sinh{\frac{\vartheta}{2}} e^{i\varphi} & \cosh{\frac{\vartheta}{2}}\\
\end{array}
\right]\\
&+\sigma_{+}\otimes\left[
\begin{array}{cc}
\cosh\frac{\theta}{2}e^{i\zeta} & -\sinh\frac{\theta}{2} e^{-i\phi} \\
-\sinh\frac{\theta}{2} e^{i\phi} & \cosh\frac{\theta}{2}e^{i\zeta}\\
\end{array}
\right]\\
&+\sigma_{-}\otimes\left[
\begin{array}{cc}
\cosh\frac{\theta}{2}e^{-i\zeta} & -\sinh\frac{\theta}{2} e^{-i\phi} \\
-\sinh\frac{\theta}{2} e^{i\phi} & \cosh\frac{\theta}{2}e^{-i\zeta}\\
\end{array}
\right],\\
\end{split}\label{Eq4}
\end{equation}

where the matrix $U$ enables $h_{0}$=$\ell$$\cosh{\vartheta}$, $f_{1}(\boldsymbol{k})$=$\ell$$\sinh{\vartheta}e^{-i\varphi}$, $f_{2}(\boldsymbol{k})$= $l\sinh{\theta}e^{-i\phi}$and $g_{\chi}(\boldsymbol{k})$=$l\cosh{\theta}e^{i\zeta}$. This para-unitary transformation in the BdG framework provides a robust mechanism to study stability and dynamics of  topological magnons with chirality. We denote the left (right)-hand chirality of magnon bands from $\chi$=$\langle0|\Psi_{k}\mathbf{S}^{z}\Psi_{k}^{\dag}|0\rangle$=$\pm 1$ satisfying the generalized orthonormal condition $\langle\langle\Psi_{m}|\Psi_{n}\rangle\rangle$=$\Psi_{m}^{\dag}(\sigma_{0}\otimes\tau_{z})$ $\Psi_{n}$. Taking into account this combined symmetry, the Berry connection is redefined by the BdG inner product of the transformed quasi-particle states

\begin{align}
\mathbf{A}_{n}=i[(\sigma_{0}\otimes\tau_{z})U^{\dag}(\sigma_{0}\otimes \tau_{z})\nabla U]_{nn}, \label{Eq5}
\end{align}

where the Berry curvature $\Omega_{n}$=$\nabla\times\mathbf{A}_{n}$ is derived into $i[(\sigma_{0}\otimes\tau_{z})\nabla U^{\dag}$$\times$$(\sigma_{0}\otimes \tau_{z})\nabla U]_{nn}$. Using the hyperbolic parameterization $l$ and $\ell$ for Eq.~(\ref{Eq4}), we rewrite the energy spectra in the form of $\hbar\omega_{\chi}^{\pm}$= $\sqrt{l^2+\ell^2 \pm 2l\ell(\cosh{\theta}\cosh{\vartheta}+\sinh{\theta}\sinh{\vartheta}\cos{(\phi-\varphi}))}$. As shown in Fig.~\ref{Fig2}, topological dynamics of altermagnetic magnon $|\Psi_{n}\rangle$ can be obtained via Berry curvature

\begin{equation}
\begin{split}
\Omega_{n}=&-\mathrm{Im}\big\langle\nabla\Psi_{k}|\times(\sigma_{0}\otimes \tau_{z})|\nabla\Psi_{k}\big\rangle\\
=&\frac{\sinh{\theta}}{2}\nabla\theta\times\nabla\phi+\frac{\sinh{\vartheta}}{2}\nabla\vartheta\times(\nabla\varphi+\nabla\zeta),
\end{split}\label{Eq6}
\end{equation} 

where the $\Omega_{n}$ along with $\mathbf{A}_{n}$ introduces a $\mathbb{Z}_{2}$ topological invariant to the magnon bands. The Kramers degeneracy form nodal-lines centeredat the time-reversal-invariant momenta (TRIM) points $\Gamma_{i}$ as shown in Fig.~\ref{Fig2}. Accompanied by nonrelativistic spin splittings with a common spin-quantization axis, $\mathbf{S}^{z}$ remains a good quantum number in the absence of SOC. We define ${\nu_n}$ as a practical form of $\mathbb{Z}_2$ invariant for the $n$-th Kramers pair of bands

\begin{equation}
\begin{split}
\nu_{n}&=\frac{1}{2\pi}\big[\Upsilon_{n}-\iint_{EBZ} dk_x dk_y\Omega_{n}\big] 
\quad {mod}\quad2,\\ 
\end{split}\label{Eq7}
\end{equation}

where $\Upsilon_{n}$ is the Berry phase calculated by the Wilson loop method \cite{MHZhang3}, and $EBZ$ stands for the eﬀective Brillouin zone. Numerical estimate shows that ${\nu_n}=1$ yielding a nonzero $\mathbb{Z}_2$ invariant to confirm the topological protection of chiral nodal-lines.


\begin{figure}[t]
\centering
{
\includegraphics[width=\linewidth]{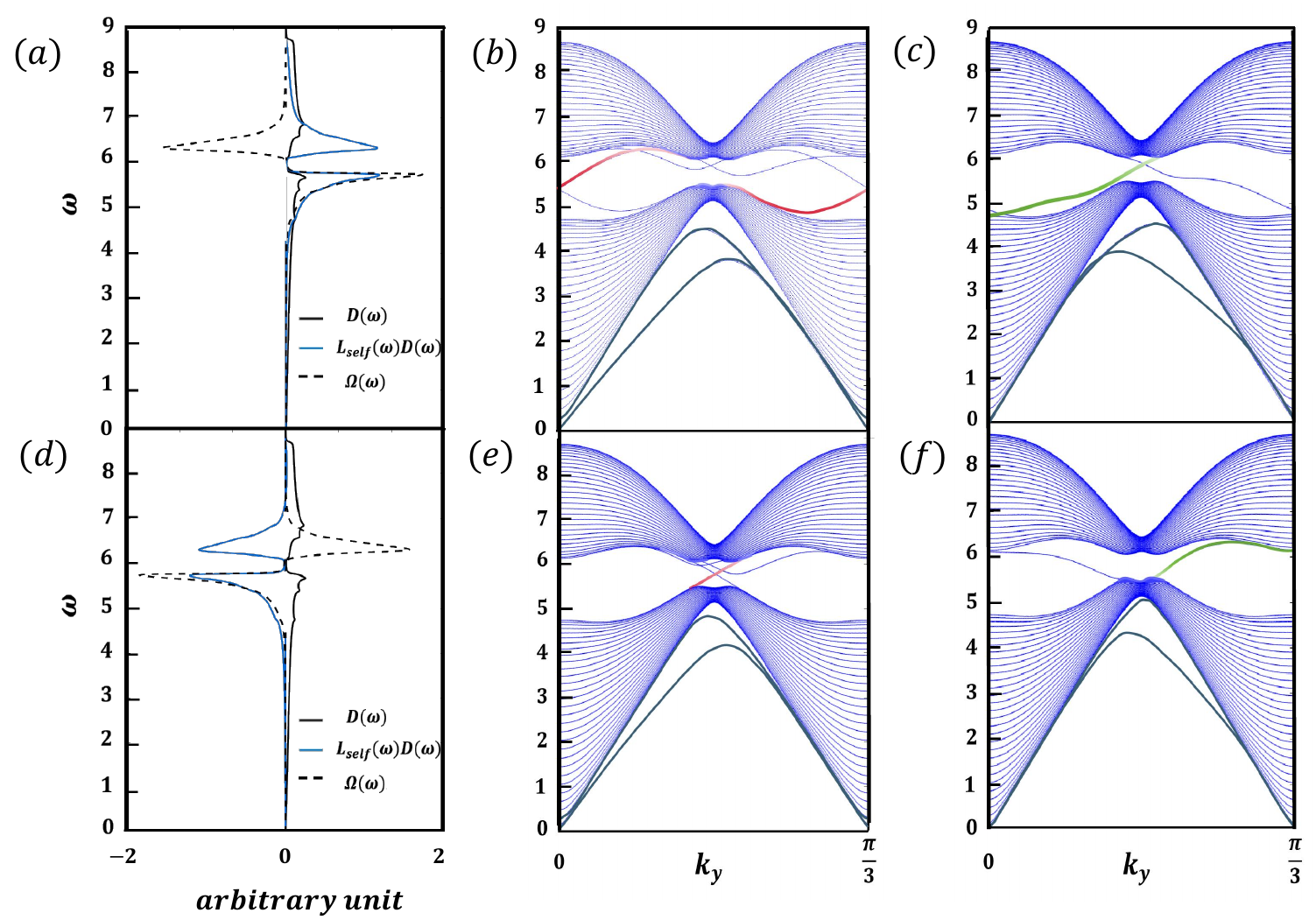}
}
\caption{Weyl magnons. (a), (b), (c) are the chiral edge modes, DOS $D(\omega)$, product density $\Omega(\omega)D(\omega)$ and angular momentum product density $L_{self}(\omega)D(\omega)$ for $J_{2}'$=0.8, $\delta$=0.4 and $J_{2}''$=0.2 respectively. (d), (e), (f) are the same quantities for $J_{2}'$=0.8, $\delta$=$-$0.4 and $J_{2}''$=0.2 . The insets show the magnon Berry curvature at the two Weyl points.}
\label{Fig3}
\end{figure}

\textit{\color{blue} Weyl and Nodal-line Magnons.-} Despite the breaking of $\mathcal{PT}$ symmetry, the nonrelativistic symmetry combines $\mathcal{T}$ with mirror $\mathcal{M}$ symmetry conserving the altermagnetic honeycomb monolayers. We define the inversion operator $\mathbf{R}$ as $\sigma_{0}\otimes$($\tau_{+}+\tau_{-}$) which interchanges the different sublattices in each chirality respectively. Although the altermagnetic nodal-lines are symmetry protected, we can further confirm their topological protection from the parity eigenvalues at $\Gamma_{i}=\{ \mathit{\Gamma} , X, Y, M \}$. The eigenvectors of  $\sigma_{0}\otimes\tau_{z}$$H_{\boldsymbol{k}}$ are $\Psi^{\dag}(\Gamma_{i})$=$[\cosh{\frac{\theta}{2}(\Gamma_{i})}, -\sinh{\frac{\theta}{2}(\Gamma_{i})}e^{-i\phi(\Gamma_{i})},$ $-\cosh{\frac{\vartheta}{2}(\Gamma_{i})}e^{i\chi}, -\sinh{\frac{\vartheta}{2}(\Gamma_{i})}e^{-i\varphi(\Gamma_{i})}]$, where the preservation of mirror symmetry enforces $\pm$1 parity eigenvalues for the states at the TRIM points. Satisfying $\mathbf{R}UH_{\boldsymbol{k}}$-$UH_{\boldsymbol{-k}}\mathbf{R}$=0, the parity eigenvalues of $\mathbf{R}$ are $\xi_{n}(\Gamma_{i})$ at $\Gamma_{i}$ associated with the nodal-line magnons

\begin{equation}
\begin{split}
(-1)^{\nu}=\prod_{n}\prod_{i}\xi_{n}(\Gamma_{i}),\\
\end{split}
\end{equation}

where the $\mathbb{Z}_2$=1 characterizes topological state via $\nu$ index confirming the odd number of chiral nodal-lines (see Supplemental Materials \cite{MHZhang3}). We introduce the pseudo-time-reversal operator $\mathcal{T}$ =$\sigma_{z}\otimes$i$\tau_{y}\mathcal{K}$, where $\mathcal{K}$ denotes the complex conjugate operator. We further analyze the wavefunctions with the six-fold rotational $C_{6v}$ operation generated anisotropic g or i-wave forms. The sublattices are locally equivalent but globally inequivalent with the presence of 2-fold spin rotation symmetry $C_{2z}$ in the altermagnetic wallpaper group p$\bar{6}m$ as shown in Fig.~\ref{Fig3}. In the hexagonal lattice monolayer, the combination of $C_{2z}$ and $C_{6v}$ symmetry and the broken $\mathcal{PT}$ symmetry impose constraints in the altermagnetic splitting. 

Since the Heisenberg equation of motion remains invariant under mirror symmetry, the energy levels degenerate along the mirror line in the magnon spectrum. For the $J'_{2}$$<$$J_{2}$ case, the nodal-lines evolve into the Weyl semimetal structure for a  finite $\delta$. In Fig.~\ref{Fig3}, pairs of Weyl points with opposite chiralities emerge at $W_{\chi}$=($\frac{\chi}{\sqrt{3}}\arccos\eta$,$\frac{\chi}{3}\arccos\eta$) with $\eta$=$\frac{-J_{1}^{2}+2h_{0}J'_{2}}{J_{1}^{2}-2h_{0}J_{2}}$ along the high symmetry axis $\Gamma$-$M$. The Weyl magnons $\hbar\omega_{\chi}^{\pm}$ are given with the related functions 

\begin{equation}
\begin{split}
\Lambda(\boldsymbol{k})&=h_{0}^{2}-f_{1}(\boldsymbol{k})f_{1}^{\dag}(\boldsymbol{k})+g_{\chi}(\boldsymbol{k})g_{\chi}^{\dag}(\boldsymbol{k})-f_{2}(\boldsymbol{k})f_{2}^{\dag}(\boldsymbol{k}),\\
\lambda(\boldsymbol{k})&=\sqrt{\lambda_{1}^{2}(\boldsymbol{k})+\lambda_{2}(\boldsymbol{k})},
\end{split}
\end{equation}
where $\lambda_{1}(\boldsymbol{k})$=$f_{1}(\boldsymbol{k})f_{2}^{\dag}(\boldsymbol{k})+f_{2}(\boldsymbol{k})f_{1}^{\dag}(\boldsymbol{k})+2h_{0}\mathrm{Re}g_{\chi}(\boldsymbol{k})$,  and $\lambda_{2}(\boldsymbol{k})$=$4\mathrm{Im}g_{\chi}^{2}(\boldsymbol{k})[h_{0}^{2}-f_{2}(\boldsymbol{k})f_{2}^{\dag}(\boldsymbol{k})]$. With the crossings residing at $\lambda_{1}(\boldsymbol{k})$=0, we construct an effective Hamiltonian by introducing the projection operator $\mathbf{P}$=$U (\sigma_{0}\otimes \tau_{z})$$(U)^{\dag}(\sigma_{0}\otimes \tau_{z})$=$(\sigma_{0}\otimes \tau_{z})U$$(\sigma_{0}\otimes \tau_{z})(U)^{\dag}$. In the vicinity of $W_{\chi}$, we expand the eigenvectors to the first order and project the Bose BdG Hamiltonian into the subspace

\begin{equation}
\begin{split}
H^{W_{\chi}}_{eff}&=\sqrt{E_{W_{\chi}}}\sigma_{0}+v_{y}q_x\sigma_{y}\pm v_{x}q_y\sigma_{x}\pm\frac{\Delta_{W_{\chi}}}{2}\sigma_{z},\\
\end{split}
\end{equation} 

where $E_{W_{\chi}}$=$h_{0}^{2}-3J_{1}^{2}-2J_{1}^{2}\eta$+ $\frac{4(J_{2}-J'_{2})^{2}J_{1}^{4}}{(J_{1}^{2}-2h_{0}J_{2})^{2}}$, $v_{x}$ and $v_{y}$ are the anisotropic velocities of magnons around the Weyl points \cite{MHZhang3}. The $J_{2}''$ term creates the mass term as an operator $\mathbf{M}$ = $\sigma_{+}^{0}\otimes$($M(\boldsymbol{k})\sigma_{0}$), where the relative momentum of magnons near the Weyl points follows a quadratic form. $M(\boldsymbol{k})$ = $2J_{2}''\big[1$ - $M_{c} \cos(\boldsymbol{k\cdot a_1}$+$\boldsymbol{k\cdot a_2})$ + $M_{s} \sin(\boldsymbol{k\cdot a_1}$ + $\boldsymbol{k\cdot a_2})\big]$ is the staggered coupling strength generating the topological gap $\Delta_{W_{\chi}}$. The general hopping of $J_{2}''$ term respects the underlying altermagnetic symmetry, ensuring that the mass terms introduced with appropriate phase factors prevent spin mixing. Thus the Berry connection is modified with $\mathbf{M}$ operator as $i[(\sigma_{0}\otimes\tau_{z})$$\mathbf{M}U^{\dag}$ $(\sigma_{0}\otimes \tau_{z})$$\nabla U]$, and the Berry curvature is approximated as $\Omega_{n}$=$\frac{\Delta_{W_{\chi}}}{2(v_{x}v_{y})^{\frac{3}{2}}}$. Contrary to the altermagnetic model based on a modified Kane-Mele framework, we observe the localization of the Berry flux around Weyl points \cite{HaldaneKaneMeleModel,modifiedKaneMele}. The monopole-type Berry curvature acts as a source or drain in the Brillouin zone, ensuring robust topological properties under perturbations. 

When the mass terms are present, the magnon dynamics are governed by the unique modulations in the exchange fields introduced by altermagnetism, where the effective mass $m^{*}$ of Weyl magnons is described by the quadratic dispersion in the vicinity of energy valleys $\hbar\omega^{\pm}_{W_{\chi}}(q)$$\approx$$\sqrt{E_{W_{\chi}}\pm c_{\omega}q^{2}}$. As shown in Fig.~\ref{Fig3}, the coupling of Berry curvature and chirality at the valley points activates the valley-resolved DOS, labelling the valley indexes $\chi$ for Weyl points in collinear altermagnets. The van Hove singularities (vHSs) enhance the magnon DOS with a logarithmic divergence, facilitating stronger magnon-magnon interactions and potential phase transitions.

We observe the annihilation of pairs of Weyl points and nodal-lines touch at the $M$ point forming a phase boundary when the condition $J'_{2}$$=$$J_{2}$ is satisfied. When $J'_{2}$$>$$J_{2}$, we find chiral nodal-line magnons by preserving the threefold symmetric band structure in momentum space. The $\delta$ term opens a finite gap, whereas the symmetry is characterized by the $\mathbb{Z}_2$ topology. Although the combined symmetry forbids the $\delta$-gapped topology to have a nonzero Chern number, the corresponding edge modes with double degeneracy denote the nontrivial topological phase with a $\delta$-induced topological gap. The $\mathbb{Z}_2$ invariance ${\nu_n}$ defines the topological protection of nodal-lines preserving Kramers degeneracy.
We effectively capture the anomalous velocity of altermagnetic quasiparticles via semiclassical equations, where chiral texture can be reflected in both real and momentum spaces. 


\begin{figure}[t]
\centering
{
\includegraphics[width=\linewidth]{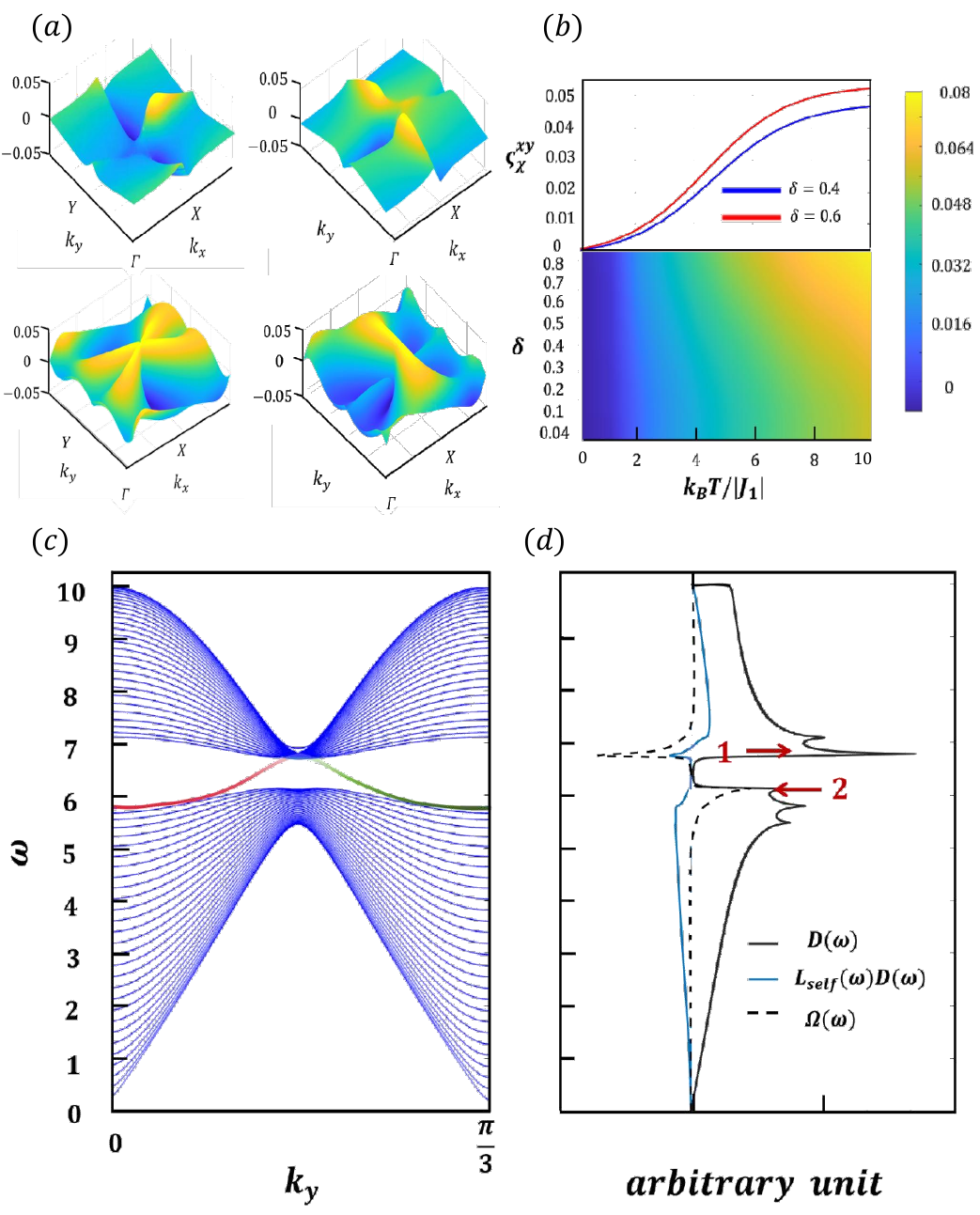}
}
\caption{Weak topological magnons for $J_{1}$=1 and $J_{2}''$=0. (a) Berry curvatures of the acoustical (optical) magnon bands with different chiralities. (b) The magnon spin Nernst conductivity as a distribution of temperature and $\delta$. (c) Topological edge modes of the finite ribbon geometry for $J_{2}'$=1.2 and $\delta$=0.4 . (d) The $D(\omega)$, $\Omega(\omega)D(\omega)$ and $L_{self}(\omega)D(\omega)$ for the right-hand chirality, where the two von Hove singularities are marked.}
\label{Fig4}
\end{figure}

\textit{\color{blue} Topological Responses.} The non-relativistic nature of band structures ensures protection through symmetries, fundamentally altering the way that the spin-polarized currents generate and propagate. Spin accumulations originated from this altermagnetic spin splitting induce an unequal distribution of spin states with difference in populations. We explore the transport characteristics when a temperature gradient $\nabla T$ is applied along different edge terminations, focusing on the anisotropic spin current $\boldsymbol{J}^{T}_{\chi}$=$\varsigma_{\chi}\mathbf{z}\times\nabla T$. Integrating over momentum and incorporating both the band topology and Berry curvature effects, the completely decoupled chiral magnons aid in the stability and robustness of the induced spin currents
\begin{equation}
\begin{split}
\boldsymbol{J}^{T}_{\chi}=&\mathbf{z}\times\nabla T \sum_{n,\boldsymbol{k}} c_{1}[n_{B}(\boldsymbol{k})]\chi\Omega_{n}(\boldsymbol{k}),\\
\end{split}
\end{equation} 

where $c_1[x]$=$(1+x)\ln(1+x)$-$x\ln x$, and $n_{B}(\boldsymbol{k})$ is the Bose-Einstein distribution. The specific spin configurations impact the Hall conductivity $\varsigma_{\chi}$=$\sum_{n}\int d\omega$$c_{1}[n_{B}(\omega)]\chi\Omega_{n}(\omega)$ displaying a spin signal difference detectable along the transverse direction through spin accumulation measurements \cite{MHZhang3}. The inherent imbalance in magnon transport enables the anisotropic magnon Hall effect that is robust against impurities and defects, where the zigzag edges tend to support chiral edge states characterized by nonlinear dispersion relations. The Weyl magnons in the armchair terminated edges project onto the same point, hosting pairs of degenerate chiral states.

As the longitudinal structure can be either zigzag or armchair, a discernible difference in spin signals will manifest along the transverse direction. Characterized by their ability to propagate without backscattering, the edge modes possess accumulations of topological angular momentum where the left- and right-hand chiral magnons segregate perpendicularly to the applied stimulus. We derive the response of magnon angular momentum via the particle self-rotation operator and current operator $\boldsymbol{L}$=$\boldsymbol{L}_{self}$+$\boldsymbol{L}_{edge}$ \cite{Murakami1,Murakami2}


\begin{equation}
\begin{split}
\boldsymbol{L}_{self} &= m^{*}\mathrm{Im} \sum_{n,\boldsymbol{k}} \big \langle\langle \nabla\Psi_{k}|\frac{n_{B}} {2k_B} (\omega_{n \boldsymbol{k}} -H)\chi|\nabla\Psi_{k}\big \rangle\rangle,\\
\boldsymbol{L}_{edge} &= m^{*}\mathrm{Im} \sum_{n,\boldsymbol{k}} \big \langle\langle \nabla\Psi_{k} |\big[Tc_1 (n_{B}) -\frac{n_{B} \omega_{n \boldsymbol{k}} }{k_B}\big]\chi|\nabla\Psi_{k}\big \rangle\rangle.\\ 
\end{split}
\end{equation} 

The angular momentum textures of topological magnons manifest as mechanical rotations when the system undergoes changes in temperature or magnetic field, revealing the intrinsic link between the band topology and macroscopic mechanical effects. We schematically highlight the high-energy edge states in Fig.~\ref{Fig3} and ~\ref{Fig4}, which are produced by the Berry curvature as an equivalent magnetic field in momentum space. When the thermal agitations are negligible, the spin-up and -down magnons are driven unequally via nonzero $\mathbf{M}$ operator. While magnons deflect perpendicularly to the gradient due to the Berry curvature, the differential gyromagnetic ratios $\gamma_{self}^{\ast}$=$\frac{\partial \boldsymbol{L}_{self}/\partial T}{\partial (S-m)/\partial T}$ and $\gamma_{edge}^{\ast}$=$\frac{\partial \boldsymbol{L}_{edge}/\partial T}{\partial (S-m)/\partial T}$ outline the expected directional differences in angular momentum. A compensatory mechanical rotation offsets the differential gyromagnetic ratio in the EdH effect \cite{niexin2}, providing a sensitive means to detect the chiral splitting induced by the altermagnetism.

\begin{figure}[t]
\centering
{
\includegraphics[width=\linewidth]{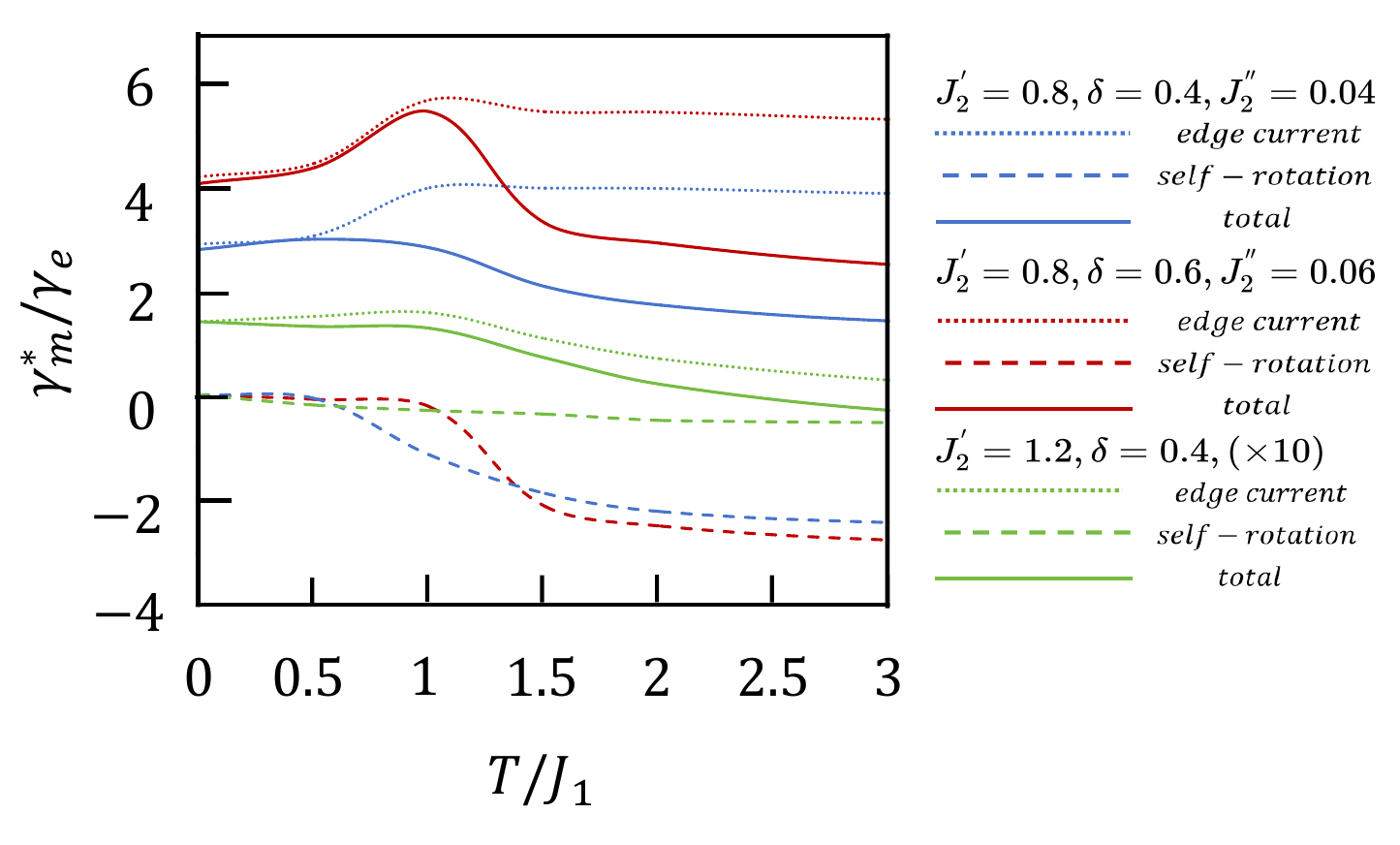}
}
\caption{Temperature dependence of the differential gyromagnetic ratio for the altermagnetic magnons with $J_{1}$=1 and $J_{2}$=0.8. Parameter choices are $\delta$=0.4, 0.6 for $J_{2}''$=0.04, 0.06 when $J_{2}'$=0.8 respectively, and $\delta$=0.4 for weak topological phase as $J_{2}'$=1.2 .}
\label{Fig5}
\end{figure}

Numerical calculations are shown in Fig.~\ref{Fig5}, demonstrating the relation between the magnon angular momentum and the magnetic moment of altermagnetic spin textures. The net differential gyromagnetic ratios of magnons are finite at the zero temperature limit, stemming from the self-rotational motions protected by the bulk topology of magnon wave packets. We investigate the unidirectional magnon transport revealing that the high temperature response is dominated by the edge currents to counterbalance the total gyromagnetic ratio $\gamma_{total}^{\ast}$=$\gamma_{self}^{\ast}$+$\gamma_{edge}^{\ast}$. By accumulating topological angular momentum, the Fig.~\ref{Fig5} shows that the magnon EdH effect manifests itself as a peak response in the optimal temperature zone. Despite the relatively weak EdH response when $J'_{2}$$>$$J_{2}$, the double degeneracy of edge modes can induce the higher-order topological phases such as corner states. 

\textit{\color{blue}Conclusion.-} We have proposed a fundamental framework of topological magnons with chirality splitting in a honeycomb altermagnet, where the symmetry breaking of the second nearest neighbor exchange couplings is used to produce Weyl and nodal-line magnons. The presence of bulk Berry curvature induces nontrivial topological response capable of generating magnon edge modes on ribbons. While the low-energy edge states share the same nature with a honeycomb antiferromagnet, the interplay between the collinear altermagnets and nontrivial bulk topology contributes to the emergence of magnon currents within the gap between the acoustic and optical magnon bands. Momentum-resolved spectroscopy can reveal the lifting of Kramers degeneracy due to the broken pseudo-time-reversal $\mathcal{T}$ symmetry manifesting as spin-split bands found in altermagnets \cite{MnTe_1}. The absence of significant second harmonic generation signals can indicate the preservation of space-inversion $\mathcal{P}$ symmetry, contributing to the appearance of non-degenerate chiral magnons\cite{MnTe,AlterM2}.

Distinguished from the DMI or other SOC induced topological magnons, altermagnetic magnons dependent on the intrinsic magnetic properties and specific lattice symmetries can induce the high-energy chiral edge transport. The chiral edge modes make the altermagnets efficient magnon generators with chiralities \cite{QuantumCircuits1}. These altermagnetic magnons can exhibit nonzero magnon spin Nernst effect. Furthermore, we calculated the angular momentum and showed the EdH effect for the altermagnetic magnons. It will be interesting to study the high-order topological phases in this model system. Besides the solid state materials, the model system can be more easily implemented in atomic spins on surface using STM, ultracold atoms in optical lattice, Rydberg atoms, and topolectrical circuits since it does not require SOC \cite{Rydberg,Coldatom,TopolectricalCircuits1}. Our study provides a new way to realize altermagnetic magnons and works as a building block for more sophisticated magnonic devices. 


\textit{\color{blue}Acknowledgements.-}
We thank Zi-Jian Xiong, Zenan Liu for helpful discussions. This project is supported by NKRDPC-2022YFA1402802, NSFC-92165204, NSFC-12494591, Leading Talent Program of Guangdong Special Projects (201626003), Guangdong Provincial Key Laboratory of Magnetoelectric Physics and Devices (No. 2022B1212010008), Research Center for Magnetoelectric Physics of Guangdong Province (2024B0303390001), Guangdong Provincial Quantum Science Strategic Initiative (GDZX2401010), and Shenzhen International Quantum Academy.


\bibliography{cite}

\end{document}


\title{Supplemental Materials for ``Topological magnons in a collinear altermagnet"}

\author{Meng-Han Zhang}
\affiliation{Guangdong Provincial Key Laboratory of Magnetoelectric Physics and Devices, State Key Laboratory of Optoelectronic Materials and Technologies, Center for Neutron Science and Technology, School of Physics, Sun Yat-sen University, Guangzhou, 510275, China}

\author{Lu Xiao}
\affiliation{Guangdong Provincial Key Laboratory of Magnetoelectric Physics and Devices, State Key Laboratory of Optoelectronic Materials and Technologies, Center for Neutron Science and Technology, School of Physics, Sun Yat-sen University, Guangzhou, 510275, China}

\author{Dao-Xin Yao}
\email{yaodaox@mail.sysu.edu.cn}
\affiliation{Guangdong Provincial Key Laboratory of Magnetoelectric Physics and Devices, State Key Laboratory of Optoelectronic Materials and Technologies, Center for Neutron Science and Technology, School of Physics, Sun Yat-sen University, Guangzhou, 510275, China}

\date{\today}
\maketitle

\section{A. Magnon Dynamics}

\begin{figure}[t]
\centering
{
\includegraphics[width=\linewidth]{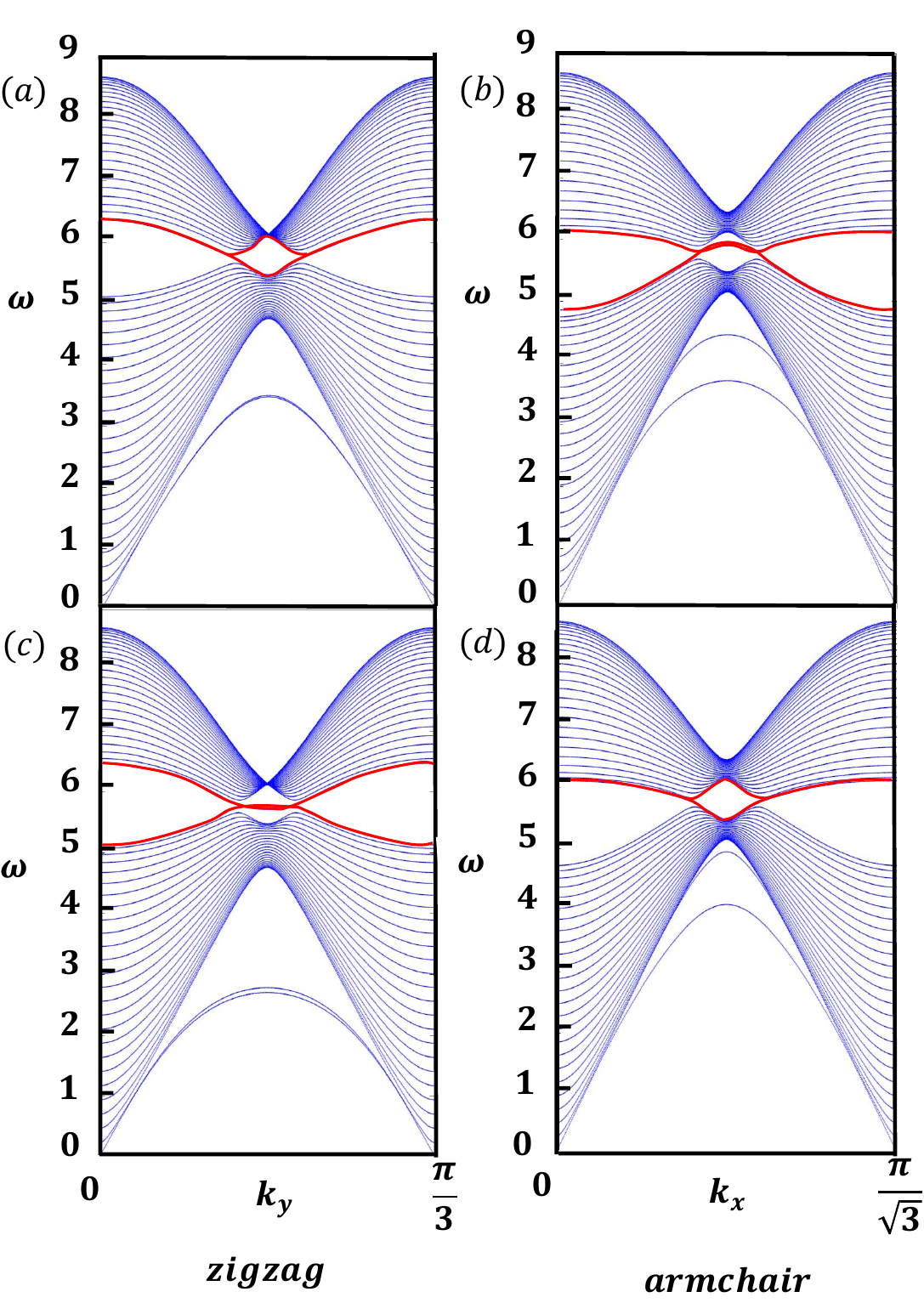}
}
\caption{Chiral splitting Weyl magnons for $J_2/J_1$=0.8, $J_2'/J_2$=0.8 and $\delta$=0.4 taking $J_1$ as the unit of energy. The (a) and (b) figures are the zigzag and armchair edge terminations for right-hand chiral magnons respectively. The (c) and (d) figures share the same edge states with left-hand chirality.}
\label{Fig1}
\end{figure}

The unit cell of the $N\acute{e}el$ ordered state in the honeycomb altermagnet consists of four different sublattices, where the primitive vectors are defined as $\boldsymbol{a_1}$=$(\frac{\sqrt{3}}{2}, \frac{-1}{2})a$ and $\boldsymbol{a_2}$=$(\frac{\sqrt{3}}{2}, \frac{1}{2})a$ in Fig.S\ref{Fig1}. We use the Holstein-Primakoff (HP) representation to study the magnetic excitations of the model and set the zigzag and armchair directions as $x-$ and $y-$axes respectively

\begin{equation}
\begin{split}
\mathbf{S}_{m\alpha }^{+}&=\sqrt{2S-\alpha^{\dag}_m\alpha_m}\alpha^{}_m, \mathbf{S}_{m\alpha }^{-}=\alpha^{\dag}_m\sqrt{2S-\alpha^{\dag}_m\alpha_m},\\
\mathbf{S}_{m\beta }^{+}&=\beta^{\dag}_m\sqrt{2S-\beta^{\dag}_m\beta_m}, \mathbf{S}_{m\beta }^{-}=\sqrt{2S-\beta^{\dag}_m\beta_m}\beta^{}_m,\\
\mathbf{S}_{m\alpha }^{z}&=S-\alpha^{\dag}_m\alpha^{ }_m, \mathbf{S}_{m\beta }^{z}=\beta^{\dag}_m\beta^{ }_m-S, 
\end{split}\tag{SA-1}
\end{equation}

where $\alpha^{\dag}_m(\alpha_m^{})$ is the bosonic magnon creation (annihilation) operator at site $m$. The original spin model can be represented by the bosonic Bogoliubov-de Gennes (BdG) Hamiltonian, leading to specific algebraic forms that are non-Hermitian yet symmetrical. We map this typical altermagnetic Hamiltonian into the Nambu basis $\psi_{k}^{\dag}$$\equiv$$[\alpha^{\dag}_{1,k}, \beta^{ }_{2,-k}, \alpha^{\dag}_{3,k}, \beta^{ }_{4,-k}]$ following the Fourier transformation:
\begin{align}
\alpha^{\dag}_{\boldsymbol{k}}=\frac{1}{\sqrt{N}}\sum_me^{i\boldsymbol{k}\cdot\boldsymbol{R}_m}\alpha^{\dag}_m.\tag{SA-2}
\end{align}

The non-Hermitian nature means that the eigenvalue problem involves solving a grand-dynamical matrix $\mathcal{H}(\boldsymbol{k})$$\equiv$$(\sigma_{0}\otimes\tau_{z})$$\mathcal{D}(\boldsymbol{k})$, which is constructed from $A(\boldsymbol{k})$ and $B(\boldsymbol{k})$\cite{DiagonalizationBdG1978}

\begin{align}
\mathcal{D}(\boldsymbol{k})&=S\left[
\begin{array}{cc}
A_{\boldsymbol{k}} & B_{\boldsymbol{k}}\\
B_{\boldsymbol{k}}^{\dag} & A_{\boldsymbol{k}} \\
\end{array}
\right], \tag{SA-4}
\end{align}
where $A(\boldsymbol{k})$ is Hermitian and $B(\boldsymbol{k})$ is normal
\begin{align}
A(\boldsymbol{k})=\left[
\begin{array}{cc}
h_{0} & f_{1}(\boldsymbol{k})\\
f_{1}^{\dag}(\boldsymbol{k}) & h_{0} \\
\end{array}
\right],\tag{SA-5} \\
B(\boldsymbol{k})=\left[
\begin{array}{cc}
g_{\chi}(\boldsymbol{k}) & f_{2}(\boldsymbol{k})\\
f_{2}(\boldsymbol{k}) & g_{\chi}(\boldsymbol{k}) \\
\end{array}
\right]. 
\tag{SA-6}
\end{align}

The $\mathcal{D}(\boldsymbol{k})$ is particularly relevant in understanding the BdG Hamiltonian's coefficients as it directly arises from the interactions within the altermagnetic lattice. Thus, $h_{0}$=$3J_{1}$+$2J_{2}$+$2J_{2}'$, $f_{1}(\boldsymbol{k})$ =$J_{1}e^{i\boldsymbol{k\cdot(-a_1+a_2)}}$  and $f_{2}(\boldsymbol{k})$=$J_{1}e^{-i\boldsymbol{k\cdot a_1}}$$+$$J_{1}e^{i\boldsymbol{k\cdot a_2}}$. The chiral splitting magnon bands are distinguished by $g_{\chi}(\boldsymbol{k})$ with $\chi$=$\pm1$. We perform $g_{1}(\boldsymbol{k})$=$2J_{2}$ $\cos(-\boldsymbol{k\cdot a_1}$$+$$\boldsymbol{2k\cdot a_2})$$+$2$J_{2}'\cos(\boldsymbol{2k\cdot a_1}$$-$$\boldsymbol{k\cdot a_2})$$-$$2i\delta$ $\sin(\boldsymbol{2k\cdot a_1}$$-$$\boldsymbol{k\cdot a_2)}$ and $g_{-1}(\boldsymbol{k})$=$2J_{2}\cos(\boldsymbol{2k\cdot a_1}$$-$$\boldsymbol{k\cdot a_2)}$ $+$$2J_{2}'\cos(-\boldsymbol{k\cdot a_1}$$+$$\boldsymbol{2k\cdot a_2})$$-$$2i\delta$$\sin(-\boldsymbol{k\cdot a_1}$$+$$\boldsymbol{2k\cdot a_2})$.

\begin{figure*}[t]
\centering
{
\includegraphics[width=\linewidth]{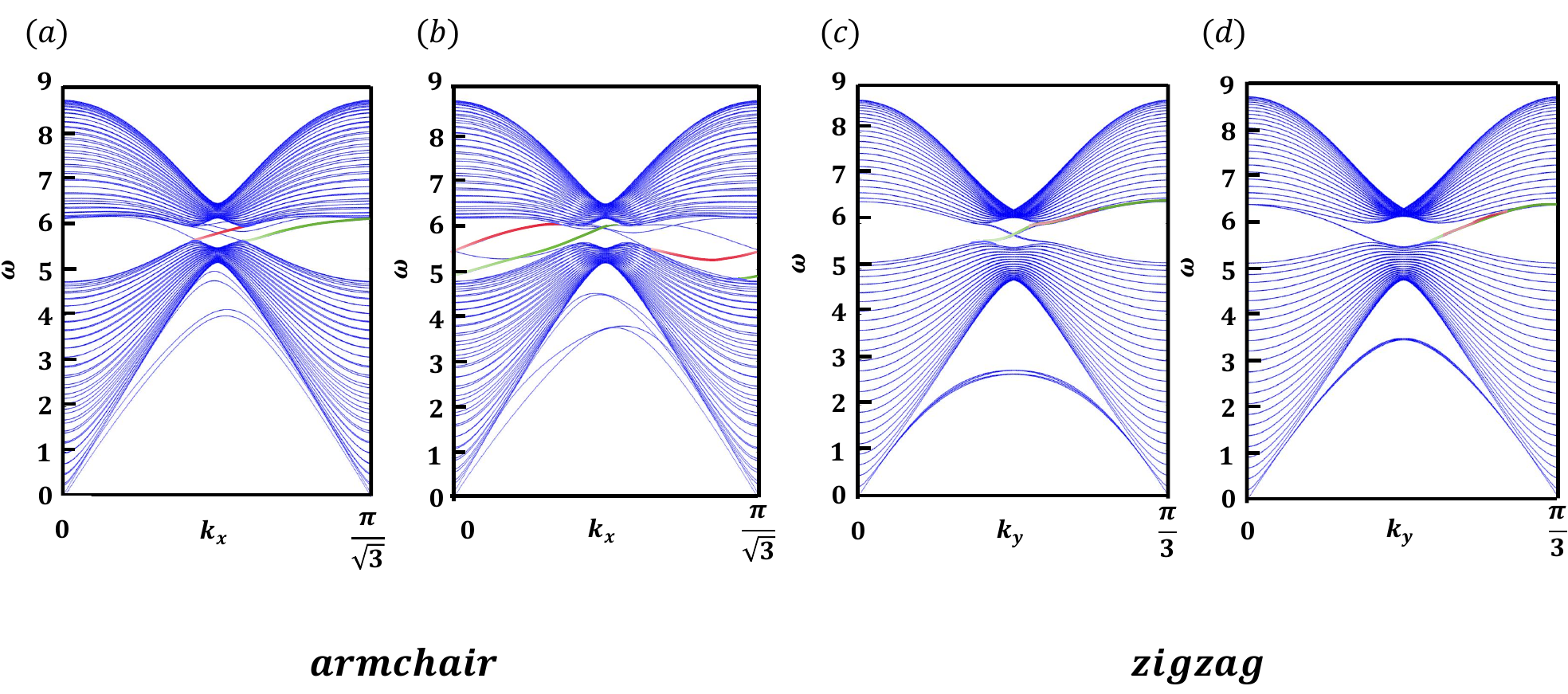}
}
\caption{The anisotropic edge modes for $M_{s}$=0.2 where other parameters are the same as Fig.S\ref{Fig1}. The (a) and (b) correspond to the chiral edge states in armchair nanoribbons, which are sensitive to magnon chirality. The (c) and (d) figures are zigzag edges enhancing the degeneracy and contributing to the overall topological properties in finite geometry.}
\label{Fig2}
\end{figure*}

In addressing the altermagnetism characterized by a quadratic bosonic Hamiltonian, we consider the grand-dynamical matrix $\mathcal{D}(\boldsymbol{k})$ that is positive-semidefinite, ensuring that all its paravalues are non-negative to support physical stability. Then, we study the dynamical matrix $D(\boldsymbol{k})$$\equiv$$A^{2}(\boldsymbol{k})-B^{\dag}(\boldsymbol{k})B(\boldsymbol{k})$ that is generally derived from the Hessian of the potential energy function. The order of the $D(\boldsymbol{k})$ is typically half of the order of $\mathcal{D}(\boldsymbol{k})$, where the eigenvalues related to the squares of the mode energies can also be derived from the paravalues of $\mathcal{D}(\boldsymbol{k})$. Providing insights into the magnon dynamics in the reciprocal space, both $D(\boldsymbol{k})$ and $\mathcal{D}(\boldsymbol{k})$ prevent imaginary frequencies which indicate dynamic instabilities. 

The dynamical matrix $D(\boldsymbol{k})$ allows us to find the coefficients in a corresponding diagonalizing Bogoliubov transformation $U$ from its eigenvectors. Subsequently, we start from the column $\mu_{\alpha}$ of such a para-unitary matrix $U$ forming an orthogonal set of simultaneous eigenvectors of $A(\boldsymbol{k})$ and $B(\boldsymbol{k})$

\begin{align}
\Psi_{1,\alpha}\equiv C_{1,\alpha}\left[
\begin{array}{cc}
(A+\omega_{\alpha} I)\mu_{\alpha}\\ -B^{\dag}\mu_{\alpha} \\
\end{array}
\right]\equiv\left[
\begin{array}{cc}
U_{1,\alpha}\\ V_{1,\alpha} \\
\end{array}
\right],\tag{SA-7} \\
\Psi_{3,\alpha}\equiv C_{3,\alpha}\left[
\begin{array}{cc}
-B\mu_{\alpha}\\ (A+\omega_{\alpha} I)\mu_{\alpha} \\
\end{array}
\right]\equiv\left[
\begin{array}{cc}
U_{3,\alpha}\\ V_{3,\alpha} \\
\end{array}
\right], \tag{SA-8}
\end{align}

where $\omega^{2}_{\alpha}$ is the corresponding eigenvalue, and the same processing method is applied for $\beta$ mode. We calculate the positive-semidefinite $\mathcal{D}(\boldsymbol{k})$ from the limit of the zero mode energies for the wave vector tending to $\Gamma$ point. Considering $\mathcal{D}(\boldsymbol{k})$ as a limiting case of positive-definite matrix leads to a direct and profound insight into the unitary diagonalization and their implications for mode energies. Rather than generalizing the procedure of the Cholesky decomposition, unitary diagonalization provides an approach that remains applicable even with certain isolated zero mode, where the relations between the elements of $U$ and the coefficients in the original hamiltonian are apparent.

Preserving the bosonic commutation relation $U^{\dag} (\sigma_{0}\otimes \tau_{z}) U$ =$\sigma_{0}\otimes \tau_{z}$, we give the general form of the BdG Hamiltonian  $\mathcal{H}(\boldsymbol{k})$
\begin{align}
\mathcal{H}(\boldsymbol{k})=\sum_{\boldsymbol{k}}\psi^{\dag}_{\boldsymbol{k}}\mathcal{D}(\boldsymbol{k})\psi^{}_{\boldsymbol{k}}, \tag{SA-9}
\end{align}

where the para-unitary matrix $U$ is introduced to diagonalize the BdG Hamiltonian 

\begin{align}
U^{\dag}\mathcal{D}(\boldsymbol{k})U =\left[
\begin{array}{cc}
E_{\boldsymbol{k}} &  \\
  & E_{-\boldsymbol{k}} \\
\end{array}
\right], \tag{SA-10}
\end{align}

where $E_{\boldsymbol{k}}$ is a diagonal matrix, and the eigenvectors obtained from the Nambu spinor represent the excitation spectrum of the quasi-particles and quasi-holes, reflecting both particle-like and hole-like states

\begin{align}
\Psi^{}_{\boldsymbol{k}}=U^{\dag}\psi^{}_{\boldsymbol{k}}. \tag{SA-11}
\end{align}

\begin{figure*}[t]
\centering
{
\includegraphics[width=\linewidth]{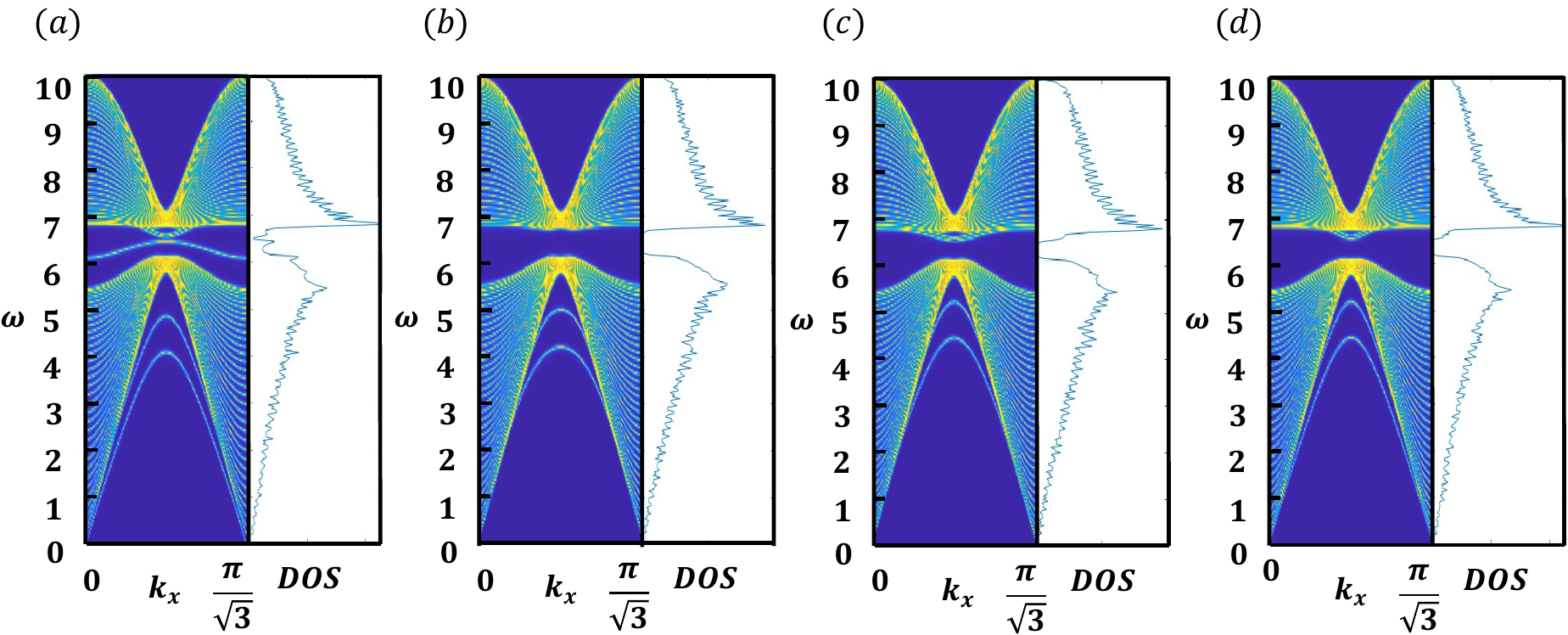}
}
\caption{ The armchair nanoribbons with $J_2/J_1=$0.8, $J_2'/J_2=$1.2 and $\delta=$0.4 are fundamentally influenced by magnon chirality. (a) For $M_{c}$=0.2 the van Hove singularities (vHs) manifest due to the double degenerate edge states with $\chi=$1. (b) The vHs disappear with the flat edge modes when $\chi=-$1. (c) Their vHs for $\chi=$1 can be altered by the reduction of $M_{c}$, or the adjustion of any other coefficient of $\mathbf{M}$ operator. (d) The $\chi=-$1 chiral armchair ribbons show behavior dependent on dopants and strain, which typically results in a logarithmic divergence of the magnon DOS.}
\label{Fig3}
\end{figure*}

During the transformation process, the Bose commutation relation should remain unchanged $[\Psi^{}_{m}(\boldsymbol{k}), \Psi^{\dag}_{n}(\boldsymbol{k'})]$=$(\sigma_{0}\otimes\tau_{z})_{mn}\delta_{\boldsymbol{k},\boldsymbol{k'}}$. The $\sigma_{0}\otimes\tau_{z}$ ensures the magnon BdG Hamiltonian remains consistent with the particle-hole symmetry\cite{SU3ChiralMagnons}.

\section{B. Magnon Topology}

Requiring careful consideration of the particle-hole symmetry, the BdG Hamiltonian solves eigenproblems of the form $(\sigma_{0}\otimes\tau_{z})$$\mathcal{D}(\boldsymbol{k})$ and interprets the resulting spectrum, where half of the eigenvalues correspond to particle-like excitations and the other half correspond to hole-like excitations. The identity projection operator $\mathbf{P}$ assists in isolating the BdG Hamiltonian's eigenstates, thereby facilitating a straightforward calculation of altermagnetic magnon energies and wavefunctions

\begin{align}
\begin{split}
\mathbf{P}&= U (\sigma_{0}\otimes \tau_{z}) U^{\dag}(\sigma_{0}\otimes \tau_{z})\\
&=(\sigma_{0}\otimes \tau_{z})U(\sigma_{0}\otimes \tau_{z}) U^{\dag},
\end{split}\tag{SB-1}
\end{align}

where $\mathbf{P}$ simplifies calculations of Berry Curvature and can be particularly useful in our honeycomb lattice with altermagtic band structures

\begin{equation}
\begin{split}
\Omega_{n}=&i[(\sigma_{0}\otimes \tau_{z})\nabla U^{\dag}(\sigma_{0}\otimes\tau_{z})U\\
&\times(\sigma_{0}\otimes \tau_{z})U^{\dag}(\sigma_{0}\otimes \tau_{z})\nabla U]_{nn},
\end{split}\tag{SB-2}
\end{equation} 

where $n$ is the band index. Due to the presence of particle-hole symmetry and non-Hermitian properties, the BdG inner product of the transformed quasi-particle states needs to be redefined. Ensuring the preservation of physical properties and the generalized orthonormal condition$\langle\langle\Psi_{m}|\Psi_{n}\rangle\rangle$=$\Psi_{m}^{\dag}(\sigma_{0}\otimes\tau_{z})$ $\Psi_{n}$, the Wilson loop method provides a more practical formulation for calculating the Berry phase

\begin{align}
\begin{split}
\Upsilon_{n}&=\mathrm{Im}\ln[ \langle\langle\Psi_{n}\mathbf{(R_{0})}|\Psi_{n} \mathbf{(R_{1})}\rangle\rangle\langle\langle\Psi_{n}\mathbf{(R_{1})}|\Psi_{n} \mathbf{(R_{2})}\rangle\rangle...\\
&\langle\langle\Psi_{n}\mathbf{(R_{N-2})} |\Psi_{n} \mathbf{(R_{N-1})}\rangle\rangle\langle\langle\Psi_{n}\mathbf{(R_{N-1})} |\Psi_{n} \mathbf{(R_{0})}\rangle\rangle], 
\end{split}\tag{SB-3}
\end{align}

where $\langle\langle\Psi_{n}|\Psi_{n}\rangle\rangle$ enables a concise form of the magnon BdG Berry connection

\begin{align}
\mathbf{A}_{n}=i(\sigma_{0}\otimes\tau_{z})_{nn}\langle\langle\Psi_{n}|\nabla\Psi_{n}\rangle\rangle. \tag{SB-4}
\end{align}

Derived from the Berry connection, the $\mathbb{Z}_2$ topological invariants for altermagnetic magnons exhibit pseudo-time-reversal symmetry along with the existence of bosonic counterparts of Kramers pairs. We extend the concept of Kramers’ pair to altermagnetic magnon system via pseudo-time-reversal operator $\mathcal{T}$ = $\sigma_{z}\otimes$i$\tau_{y}\mathcal{K}$, where $\mathcal{K}$ denotes the complex conjugate operator\cite{Magnonwindingnumber}. The $\mathbb{Z}_2$ topological invariant in altermagnetic magnons can be derived from the $\Omega_{n}$ along with $\mathbf{A}_{n}$ as a more practical form of ${\nu_n}$ for the $n$-th Kramers pair of bands


\begin{align}
\nu_{n}&=\frac{1}{2\pi}\big[\Upsilon_{n}-\iint_{EBZ} dk_x dk_y\Omega_{n}\big] \quad{mod}\quad2 ,\ \tag{SB-5}
\end{align}

where $EBZ$ stands for the eﬀective Brillouin zone. We maintain $\mathbf{S}^{z}$ as a good quantum number to precisely characterize the Chern number of non-degenerate chiral magnons, which is obtained from the Berry curvature to capture the geometric features of wavefunctions

\begin{align}
\begin{split}
\Omega_{n}=&i(\sigma_{0}\otimes\tau_{z})_{nn}\langle\langle\nabla\Psi_{n}|\times|\nabla\Psi_{n}\rangle\rangle\\
=&i(\sigma_{0}\otimes \tau_{z})_{nn}\sum_{m}(\sigma_{0}\otimes \tau_{z})_{mm}\\
&\langle\langle\nabla\Psi_{n}|\Psi_{m}\rangle\rangle\times\langle\langle\Psi_{m}|\nabla\Psi_{n}\rangle\rangle, 
\end{split}\tag{SB-6}
\end{align}

where $\langle\langle\Psi_{m}|\nabla\Psi_{n}\rangle\rangle$ =$-\langle\langle\nabla\Psi_{n}|\Psi_{m}\rangle\rangle$. The reciprocal relationships can be given as $\nabla\mathcal{D}(\boldsymbol{k})$=$[\nabla, \mathcal{D}(\boldsymbol{k})]$

\begin{equation}
\begin{split}
\langle\langle\Psi_{m}|\nabla\Psi_{n}\rangle\rangle=\frac{\langle\Psi_{m}|\nabla\mathcal{D}(\boldsymbol{k})|\Psi_{n}\rangle} {E_{n}-E_{m}},
\end{split}\tag{SB-7}
\end{equation}

where gauge dependence can cause numerical instability and inaccuracies during differentiation when the phases of the eigenstates are altered. We derive $\langle\Psi_{m}|\nabla\mathcal{D}(\boldsymbol{k})|\Psi_{n}\rangle$= $\langle\Psi_{m}|\nabla(\mathcal{D}(\boldsymbol{k})|\Psi_{n}\rangle)$$-$$\langle\Psi_{m}|\mathcal{D}(\boldsymbol{k})|\nabla\Psi_{n}\rangle$ to eliminate or reduce gauge dependence for Berry curvature calculations

\begin{equation}
\begin{split}
\Omega_{n}=&i(\sigma_{0}\otimes \tau_{z})_{nn}\sum_{m\neq n}(\sigma_{0}\otimes \tau_{z})_{mm}\\
&\frac{\langle\Psi_{n}|\nabla H_{k}|\Psi_{m}\rangle\times\langle\Psi_{m}|\nabla H_{k}|\Psi_{n}\rangle}{(E_{n}-E_{m})^{2}}.\\
\end{split}\tag{SB-8}
\end{equation} 

\begin{figure}[t]
\centering
{
\includegraphics[width=\linewidth]{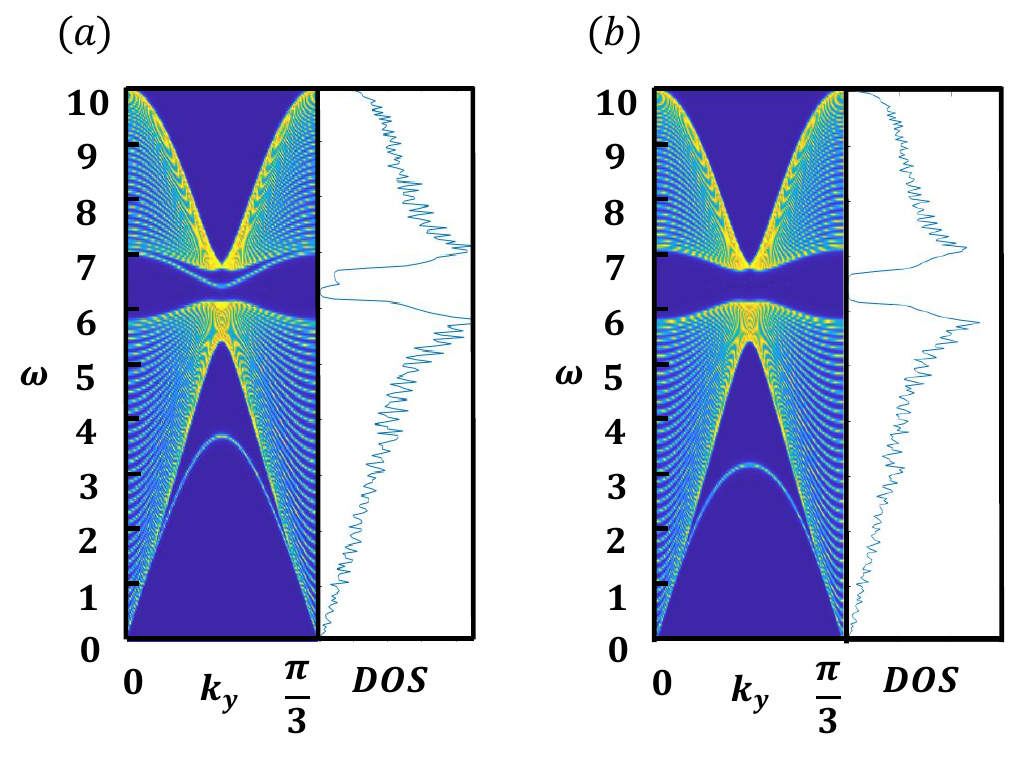}
}
\caption{Zigzag edge states exhibit topological protection and chiral-polarized magnon transport with parameter choices same as Fig.S\ref{Fig3}. (a) $M_{c}$=0.2 can lead to robust edge transport with minimal backscattering, engineering corner states driven by robust higher-order topological phases. (b) Without the $\mathbf{M}$ operator magnetic interactions induce a trivial bandgap.}
\label{Fig4}
\end{figure}

With alternating superexchange pathways, the coupling constants $\delta$ vary periodically, introducing a modulation in the magnon dispersion relation. This modulation affects the chirality of magnon modes, lifting magnon degeneracy to distinct magnon chiralities. Therefore, we rewrite our Hamiltonian in $(k_x, y)$ space along the $y$ direction
\begin{equation}
\begin{split}
\alpha^{\dag}_{ky}&=\frac{1}{\sqrt{N_{x}}}\sum_{m}e^{ik\boldsymbol{R}_{m}\cdot\boldsymbol{e}_{x}}\alpha^{\dag}_{my},\\
\beta^{\dag}_{ky}&=\frac{1}{\sqrt{N_{x}}}\sum_{m}e^{ik\boldsymbol{R}_{m}\cdot\boldsymbol{e}_{x}}\beta^{\dag}_{my}.
\end{split}\tag{SB-9}
\end{equation}

For the anisotropic edge modes, we also give the Hamiltonian for $(x, k_y)$ space
\begin{equation}
\begin{split}
\alpha^{\dag}_{kx}&=\frac{1}{\sqrt{N_{y}}}\sum_{m}e^{ik\boldsymbol{R}_{m}\cdot\boldsymbol{e}_{y}}\alpha^{\dag}_{mx},\\
\beta^{\dag}_{kx}&=\frac{1}{\sqrt{N_{y}}}\sum_{m}e^{ik\boldsymbol{R}_{m}\cdot\boldsymbol{e}_{y}}\beta^{\dag}_{mx},
\end{split}\tag{SB-10}
\end{equation}

where $y$ or $x$ runs from $i_1$ to 4($W$-1)+$i_1$ ($i_1$=$\{$1, 2, 3, 4$\}$) and $W$ denotes the number of periodic 1D chains. We can replace $k_x$ or $k_y$ by $k$. The formalism for calculating the band structure of the ribbon geometry is a $4W\times4W$ matrix-form Hamiltonian which is given by
\begin{align}
\mathcal{H}_{1D}(\boldsymbol{k})=\sum_{k}\psi^{\dag}_{k}\mathcal{D}_{1D}(\boldsymbol{k})\psi^{}_{k}, \tag{SB-11}
\end{align}
where $W$=$25$ to ensure that the results are convergent. The Hamiltonian matrix can be written as
\begin{align}
\mathcal{D}_{1D}(\boldsymbol{k})=
\left[
\begin{array}{ccccc}
\mathcal{K}_{1}(k) & \mathcal{T}(k)^{\dag} & 0 & \cdots & 0\\
\mathcal{T}(k) & \mathcal{K}(k) & \mathcal{T}(k)^{\dag} & \ddots & \vdots \\
0 & \mathcal{T}(k) & \ddots & \ddots & 0 \\
\vdots & \ddots & \ddots & \ddots & \mathcal{T}(k)^{\dag} \\
0 & \cdots & 0 & \mathcal{T}(k) & \mathcal{K}_{2}(k) \\
\end{array}
\right],\tag{SB-12}
\end{align}
where $\mathcal{T}(k)$ is transition matrix and $\mathcal{K}_(k)$, $\mathcal{K}_{1}(k)$, $\mathcal{K}_{2}(k)$ are 4$\times$4 kernel matrices. As shown in Fig.~\ref{Fig2}, the $k_{y}$-direction aligns with the zigzag orientation, while the $k_{x}$-direction is congruent with the armchair edge mode. The generalized orthonormal condition ensures that the eigenvalues obtained from diagonalizing $\mathcal{H}_{1D}(\boldsymbol{k})$ map to the poles of the Green's function

\begin{equation}
\begin{split}
\mathcal{G}^R(r, r')&=\sum_{k,n} \frac{\psi^{\dag}_{k,n}(r')\psi^{}_{k,n}(r)}{\omega+i\eta-H},\\
\mathcal{G}^A(r, r')&=[\mathcal{G}^R(r, r')]^{\dag},
\end{split}\tag{SB-13}
\end{equation}
where $\eta$ is a positive infinitesimal, $\omega$ is the excitation energy, $r$ and $r'$ represent excitation and response respectively. Reflecting the physical excitations accurately, we introduce the retarded and advanced Green's functions to calculate the transport properties of magnons. 

\section{C. Topological Phases}

In the honeycomb altermagnet, our Hamiltonian incorporates both normal and pairing terms to describe quasi-particle excitations as it reflects the preservation of particle-hole symmetry in BdG magnon system. The $\mathcal{H}_{1D}(\boldsymbol{k})$ is a central component in the study of quasi-particle excitations for systems with pairing interactions, which is particularly crucial when transforming the Hamiltonian to a form that facilitates diagonalization and analysis of its spectrum


\begin{figure}[t]
\centering
{
\includegraphics[width=\linewidth]{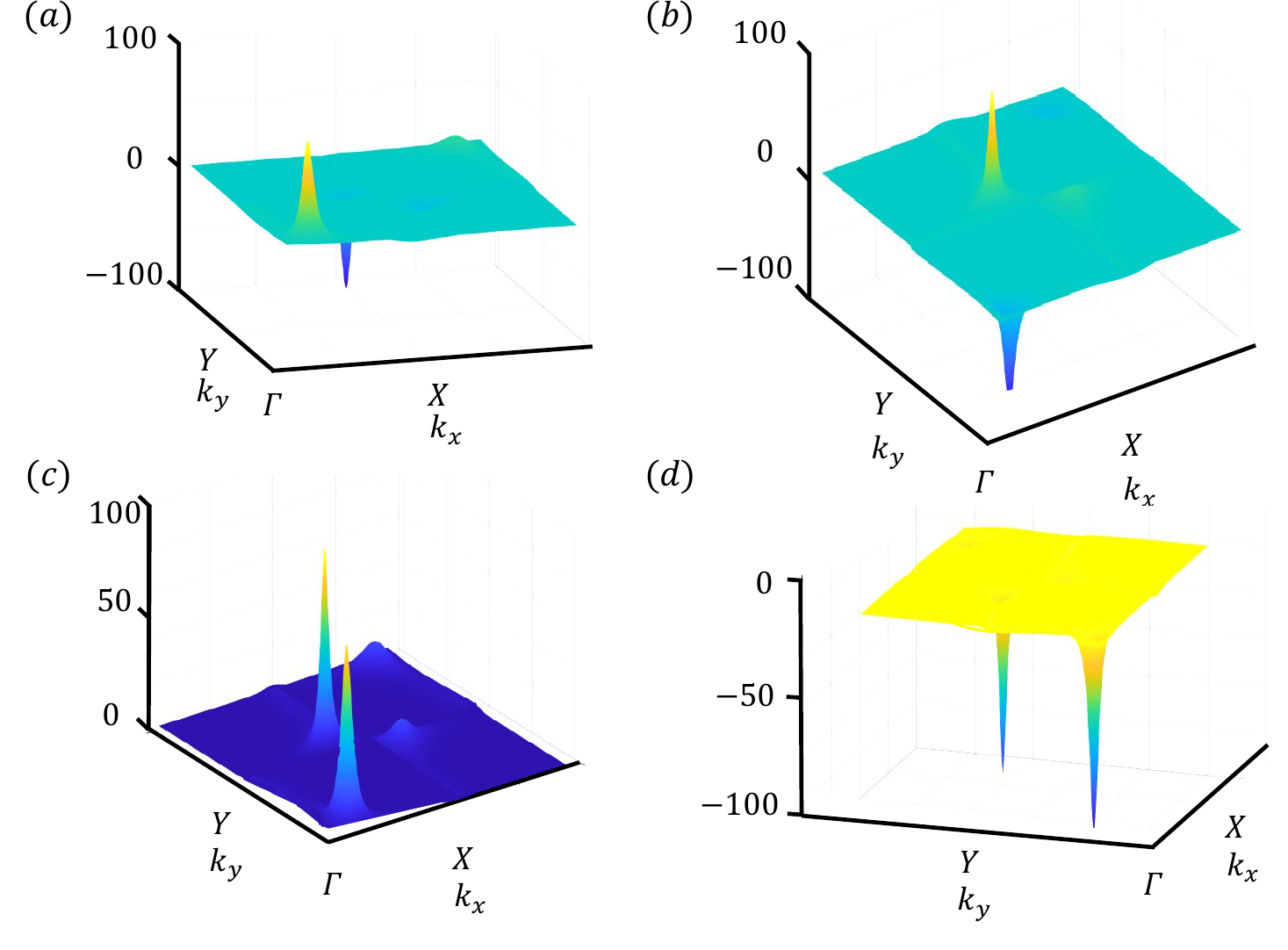}
}
\caption{The coupling of Berry curvature with chiral magnons modifies the valley Hall effect, while $M_{s}$=0.2 activates magnon valley polarization. The (a) and (b) figures respectively correspond to the acoustic and optical
branches with $M_{c}$=-0.05 and zero Chern numbers. The (c) and (d) figures are the Berry curvature for $M_{c}$=0.05 acting as an effective magnetic field in momentum space. The Chern numbers are given by $\pm1$.}
\label{Fig5}
\end{figure}

\begin{equation}
\begin{split}
A(\omega)&=\sum_{k,n} \psi^{}_{k,n}(r)\psi^{\dag}_{k,n}(r')\frac{2\eta}{(\omega-\mathcal{H}_{1D})^2+\eta^2},\\
D(\omega)&=\frac{\hbar \textmd{Tr}\big[A(\omega)\big]}{2 \pi},
\end{split}\tag{SC-1}
\end{equation}

where $A(\omega))$ is the spectral representation of the Green's function and $D(\omega)$ is the magnon density of state (DOS). As shown in Fig.~\ref{Fig4}, the van Hove singularities (vHSs) occur at points where the energy dispersion relation is stationary, which exhibit tunable topological phases. 

Since the magnon thermodynamics is tied to the the thermal fluctuations, we calculate the deviation of sublattice magnetization from the saturation value

\begin{equation}
\Delta m=S-\langle S^z_m\rangle = \langle\psi^{\dag}_m\psi_m\rangle=\sum_{n,\boldsymbol{k}} n_{B}(n,\boldsymbol{k}),\ \tag{SC-2}
\end{equation} 

where the N$\acute{e}$el temperature $T_c$ is determined by $\Delta m$($T_c$)=$S$, ensuring the validity of the semiclassical approach. Defining the sublattice magnetization $S_{\uparrow}$=-$S_{\downarrow}$= $\mathbf{S}^{z}$, we obtain the effective magnetic moment $m$=$S$-$\langle\Psi_{k}^{\dag}\Psi_{k}\rangle$=$S$-$\sum_{n,\boldsymbol{k}} n_{B}(\omega_{n \boldsymbol{k}})$. Characterized by $\chi$, the chiral nature of these bands implies that magnons exhibit directionally dependent behaviors.

Experiencing a spin-dependent Lorentz force when propagating through the collinear altermagnetic texture, magnon currents deflect in the presence of gauge fields generated by the chiral magnetic order for $J'_{2}$$>$$J_{2}$. The $\delta$ term opens a finite gap, where gap size $\Delta_{\delta}$ is $S\sqrt{\Lambda_{\delta}+ \lambda_{\delta}}$$-$$S\sqrt{\Lambda_{\delta}-\lambda_{\delta}}$ with $\Lambda_{\delta}$=$h_{0}^{2}-3J_{1}^{2}-\frac{2J_{1}^{2}}{|\eta|}$+ $\frac{4(J_{2}-J'_{2})^{2}J_{1}^{4}}{\eta^{2}(J_{1}^{2}-2h_{0}J_{2})^{2}}$ + $\frac{4\delta^{2}(\eta^{2}-1)}{\eta^{2}}$ and $\lambda_{\delta}$=$\frac{2\delta\sqrt{\eta^{2}-1}}{|\eta|}$$\sqrt{h_{0}^{2}-2\frac{J_{1}^{2}(\eta+1)}{\eta}}$. 
We derive the magnon Hall conductivity from $\sum_{n}$$c_{1}[n_{B}(\omega)]$ by the approximation as

\begin{equation}
\begin{split}
\frac{1}{e^{\frac{\sqrt{\Lambda_{\delta}+\lambda_{\delta}}}{2k_BT}}-1}-\frac{1}{e^{\frac{\sqrt{\Lambda_{\delta}-\lambda_{\delta}}}{2k_BT}}-1}\\
=\frac{\Delta^{+}-\Delta^{-}}{2\sinh{\frac{\sqrt{\Lambda_{\delta}+\lambda_{\delta}}}{2k_BT}}\sinh{\frac{\sqrt{\Lambda_{\delta}-\lambda_{\delta}}}{2k_BT}}},\\
\end{split}\tag{SC-3}
\end{equation} 

where $k_B$ is the Boltzmann constant and $T$ is the temperature. We choose the lattice constant $a$=0.1nm as the typical layer spacing for practical calculations. The $\Delta^{+}$ and $\Delta^{-}$ highlight the different topological properties of magnon BdG altermagnets

\begin{small}
\begin{equation}
\begin{split}
\Delta^{+}=&(1+\sqrt{\Lambda_{\delta}+\lambda_{\delta}})\sinh{\frac{\sqrt{\Lambda_{\delta}-\lambda_{\delta}}} {2k_BT}}\cosh{\frac{\sqrt{\Lambda_{\delta}+\lambda_{\delta}}}{2k_BT}},\\
\Delta^{-}=&(1+\sqrt{\Lambda_{\delta}-\lambda_{\delta}})\sinh{\frac{\sqrt{\Lambda_{\delta}+\lambda_{\delta}}} {2k_BT}}\cosh{\frac{\sqrt{\Lambda_{\delta}-\lambda_{\delta}}}{2k_BT}}.
\end{split}\tag{SC-4}
\end{equation} 
\end{small}

An analytic estimate of the current conductivity at low temperatures is obtained as $\varsigma_{\chi}$=$\sum_{n}\int d\omega$$c_{1}[n_{B}(\omega)]\chi\Omega_{n}(\omega)$. 

In the Weyl magnon phase, the transverse current is understood as a consequence of the $\mathbf{M}$ operator induced by the $J''_{2}$ term. Rewritten in real space$\frac{1}{2}$$\Psi_{r}^{\dag}\mathbf{M}\Psi_{r}$, the $\mathbf{M}(\boldsymbol{r})$ satisfies the commutation relation $\mathbf{M}$$\sigma_{0}\otimes \tau_{z}$$H$-$H$$\sigma_{0}\otimes \tau_{z}$$\mathbf{M}$=0. The presence of multiple Weyl point pairs resulting from the perturbed nodal line provides opportunities for manipulating and controlling their properties

\begin{small} 
\begin{equation}
\begin{split}
\lambda_{1}(\boldsymbol{k})&=2(J_{1}^{2}-6J_{1}J_{2}-4J_{2}J_{2}'-4J_{2}^{2})\cos(-\boldsymbol{k\cdot a_1}+\boldsymbol{2k\cdot a_2})\\
&+2(J_{1}^{2}-6J_{1}J_{2}'-4J_{2}J_{2}'-4J_{2}'^{2})\cos(\boldsymbol{2k\cdot a_1}-\boldsymbol{k\cdot a_2}),\\
\end{split}\tag{SC-5}
\end{equation} 
\end{small}

where the zero points of the function $\lambda_{1}(\boldsymbol{k})$ is the locations of Weyl points along the high symmetry axis $\Gamma$-$\boldsymbol{M}^{\pm}$\cite{mirrorchernbandsweyl}. By tuning the magnitude of the $J'_{2}$, we modify the behavior of the Weyl magnons. The altermagnetic spin configuration introduces non-trivial Berry curvature into the magnon bands, driving the transverse motion of magnons and leading to the topological magnon Hall effect

\begin{figure}[t]
\centering
{
\includegraphics[width=\linewidth]{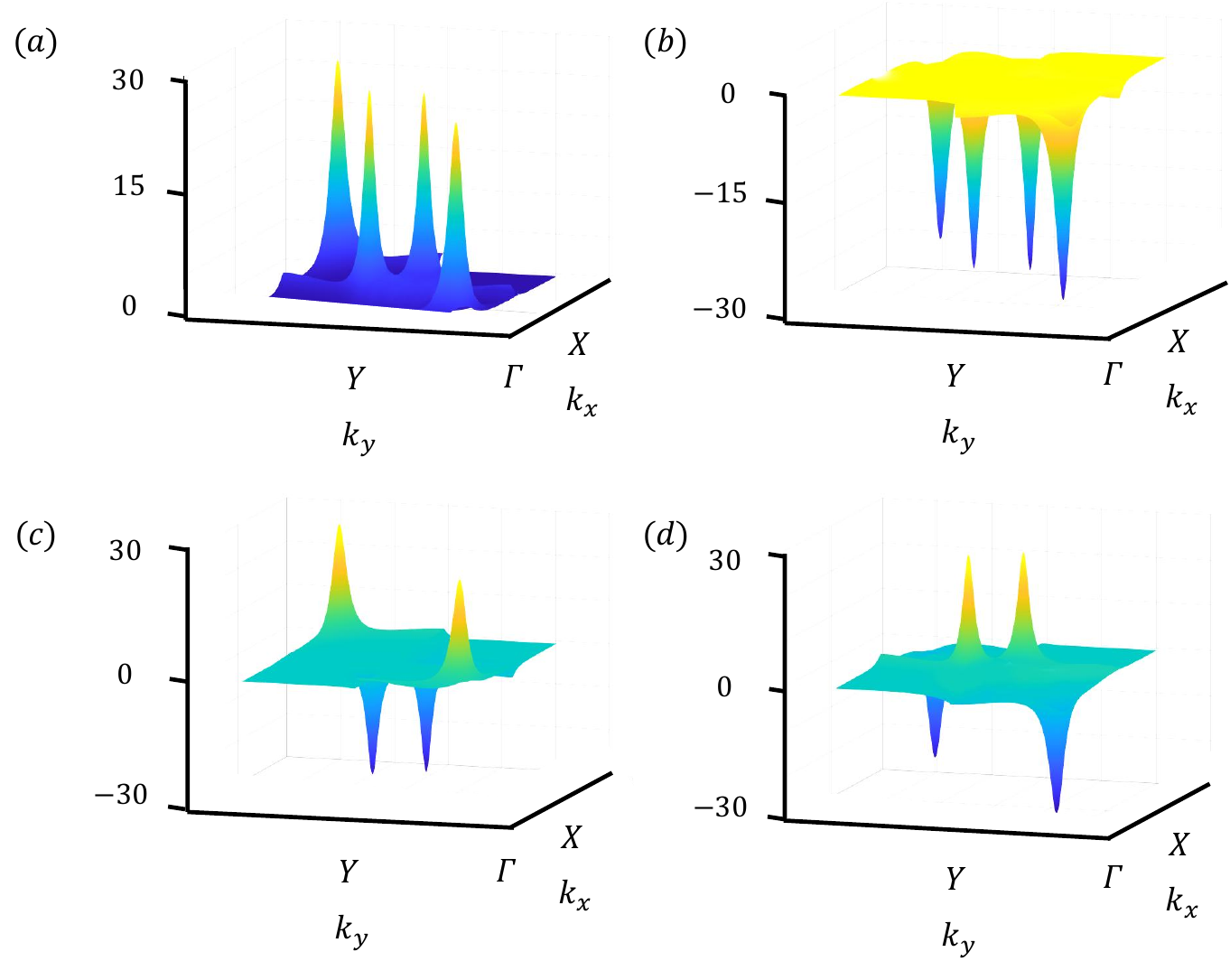}
}
\caption{Each valley possesses chiral magnonic excitations which is a measure of the direction of rotation (or handedness) of the spin waves. $M_{s}$=0.2 gives $\pm2$ Chern number for the (a) and (b) figures accordingly. The (c) and (d) figures exhibit topological trivial states with $M_{c}$=0.2 for the acoustic band and the optical band.}
\label{Fig6}
\end{figure}

\begin{equation}
\begin{split}
\Omega_{n}=&\frac{\boldsymbol{d}_{\boldsymbol{q}}\cdot[\partial_{q_{x}}\boldsymbol{d}_{\boldsymbol{q}}\times\partial_{q_{y}}\boldsymbol{d}_{\boldsymbol{q}}]}{|\boldsymbol{d}_{\boldsymbol{q}}|^{3}},
\end{split}\tag{SC-6}
\end{equation} 

where $\boldsymbol{d}_{\boldsymbol{q}}$=$v_{x}$$\boldsymbol{q_{x}}$+$v_{y}$$\boldsymbol{q_{y}}$+$\frac{\Delta_{W_{\chi}}}{2}$$\boldsymbol{q_{z}}$. In the limit of $J''_{2}$$\ll$$J_{1}$, the approximate expression of $\varsigma_{\chi}$ generated by the $\mathbf{M}$ operator of each Weyl point is given by
 
\begin{equation}
\begin{split}
\varsigma_{\chi}=&\hbar \sum_{n}\int^{\sqrt{E_{W_{\chi}}}}_{\sqrt{E_{W_{\chi}}- c_{\omega}q^{2}}} d\omega D(\omega)\chi\Omega_{\chi}(\omega)n_{B}(\omega)\\
\simeq&\frac{v_{x}v_{y}\Delta_{W_{\chi}}}{2\pi}c_{\omega} c_{\Delta_{W_{\chi}}} q^{2},
\end{split}\tag{SC-7}
\end{equation} 

where $v_{x}$=$\sqrt{3}v_{0}$, and $v_{y}$=$3v_{0}$. Our results flip sign across two von Hove singularities reflecting different chiralities with $v_{0}$= $\left[4(J_{2}J'_{2}+J_{2}^{2}\eta)-J_{1}^{2}\right]\sqrt{1-\eta^{2}}$+ $
2\sqrt{(J_{1}^{2}-h_{0}J_{2})^{2}(1-\eta^{2})+\delta^{2}(h_{0}^{2}-2J_{1}^{2}-2J_{1}^{2}\eta)}$. The nonvanishing Berry curvature causes magnon to experience an effective magnetic field, profoundly affecting the transport properties

\begin{equation}
\begin{split}
c_{\Delta_{W_{\chi}}}=\frac{\sinh{\frac{\Delta_{W_{\chi}}}{2k_BT}}}{-\cosh{\frac{\sqrt{E_{W_{\chi}}}}{k_BT}}+\cosh{\frac{\Delta_{W_{\chi}}}{2k_BT}}},
\end{split}\tag{SC-8}
\end{equation} 

where the $\delta$ dependence of the magnon spin Nernst effect has been plotted. When measured from the valley center, the relative momentum of magnons follows a quadratic dependency.

\begin{equation}
\begin{split}
\omega^{\pm}_{W_{\chi}}(q)\approx\sqrt{E_{W_{\chi}}\pm c_{\omega}q^{2}}.
\end{split}\tag{SC-9}
\end{equation} 

where $\omega^{\pm}_{W_{\chi}}(q)$ indicates a parabolic shape of the dispersion curve in the vicinity of the valley center. In the honeycomb altermagnet, the angular momentum results from the unique geometry and intrinsic magnetic interactions within the lattice. The orbital motion of a magnon wave packet is defined as the topological angular momentum without a mass term $\langle \boldsymbol{r} \times \boldsymbol{v} \rangle$


\begin{equation}
\begin{split}
L_{self} &= m^{*}\sum_{n} \int d\omega  n_{B}(\omega)\chi\boldsymbol{L}_{self}(\omega), \\
L_{edge} &= m^{*} \sum_{n} \int d\omega \big(TD(\omega)-\omega\big) n_{B}(\omega)\chi\Omega_{\chi}(\omega).\\ 
\end{split} \tag{SC-10}
\end{equation} 

where $L_{self}$ corresponds to the self-rotation of magnon wave packet and $L_{edge}$ comes from the edge current. We calculate the angular momentum of topological magnons along the integral form of $\omega$. Triggered by altermagnetic magnons, the Einstein-de Haas (EdH) effect showcases intricate angular momentum textures inherent to topological magnon bands\cite{niexin1}. We use the differential gyromagnetic ratios $\gamma_{self}^{\ast}$=$\frac{\partial \boldsymbol{L}_{self}/\partial T}{\partial (S-m)/\partial T}$ and $\gamma_{edge}^{\ast}$=$\frac{\partial \boldsymbol{L}_{edge}/\partial T}{\partial (S-m)/\partial T}$ to describe the EdH effect of topological magnons

\begin{equation}
\begin{split}
\gamma_{edge}^{\ast}=&m^{*} \frac{\sum_{n} \int d\omega \big(TD(\omega)-\omega\big) n_{B}(\omega)\chi\Omega_{\chi}(\omega)/\partial T}{\sum_{n} \int d\omega n_{B}(\omega)/\partial T},\\
\gamma_{self}^{\ast}=&m^{*} \frac{\sum_{n} \int d\omega  n_{B}(\omega)\chi\boldsymbol{L}_{self}(\omega)}{\sum_{n} \int d\omega n_{B}(\omega)/\partial T}.\\
\end{split}\tag{SC-11}
\end{equation} 

We have focused on the intrinsic magnon angular momentum accumulation to represent the pure gyromagnetic effect. The EdH effect provides an exotic bridge to link magnetism, topology, and mechanics in altermagnets. 

\bibliography{cite}